\documentclass[12pt,a4paper]{article}
\usepackage{a4,epsf,here}
\usepackage[latin1]{inputenc}
\usepackage{mathrsfs}
\usepackage{amssymb}
\usepackage{pifont,amsmath,epsfig}
\renewcommand{\i}{\ensuremath{\mathrm{i}}}      
\renewcommand{\d}{\ensuremath{\mathrm{d}}}      
\newcommand{\one}{\ensuremath{1\hspace{-0.26em}\mathrm{l}}}
\DeclareMathOperator{\sign}{\mathrm{sign}}
\DeclareMathOperator{\tr}{\mathrm{tr}}

\def\siml{\,\hbox{\kern.1em \lower.6ex \hbox{$\sim$} \kern-1.12em
          \raise.6ex \hbox{$<$} }}
\def\simg{\,\hbox{\kern.1em \lower.6ex \hbox{$\sim$} \kern-1.12em
          \raise.6ex \hbox{$>$} }}
\def\H1{\widehat{H}_1}
\def\unit{\one}
\def\M{\hbox{\large \tt M}}
\def\wkap{{\widetilde\kappa}}
\def\dmat{\hbox{\large \tt d}}
\def\bea{\begin{eqnarray}}
\def\eea{\end{eqnarray}}
\def\eq#1{(\ref{#1})}

\def\bs{\bigskip}
\def\ms{\medskip}

\def\bgrk#1{\mbox{{\boldmath $#1$ \unboldmath}}\!\!}
\newcommand{\Figurebb}[9]{
\begin{figure}[H]\begin{center}
\leavevmode
\epsfysize=#7cm
\epsfbox[#2 #3 #4 #5]{#6}
\par
\parbox{#8cm}{
\caption[figure]{\renewcommand{\baselinestretch}{0.8} \small
                                           \hspace{-0.3truecm}#9}
\label{#1}}\end{center}
\end{figure}
}

\begin{document}

\baselineskip 14pt

\centerline{\bf \Large Semiclassical trace formulae for systems} 

\ms

\centerline{\bf \Large with spin-orbit interactions:}

\ms

\centerline{\bf \Large successes and limitations of present approaches}

\bs
\bs
\bs

\centerline{\bf \large Ch Amann and M Brack}

\bs

\centerline{\small Institute for Theoretical Physics, University of
Regensburg, D-93040 Regensburg, Germany}

\bs
\bs

\centerline{\bf \large Abstract}

\bs

\baselineskip 10pt
\small{
\noindent
We discuss the semiclassical approaches for describing systems with
spin-orbit interactions by Little\-john and Flynn (1991, 1992), Frisk
and Guhr (1993), and by Bolte and Keppeler (1998, 1999). We use these 
methods to derive trace formulae for several two- and three-dimensional 
model systems, and exhibit their successes and limit\-ations. We discuss, 
in particular, also the mode conversion problem that arises in the 
strong-coup\-ling limit.
}

\bs

\centerline{v3, in print for J.\ Phys.\ A {\bf 35}, 6009 (2002)}

\baselineskip 14pt

\section{Introduction}

The periodic orbit theory (POT) initiated by M. Gutzwiller over three 
decades ago \cite{gutz} has proven to be a successful tool for the 
semiclassical description of chaotic systems \cite{pot,gubu,chao,haak}. 
Several extensions of Gutzwiller's semiclassical trace formula to systems 
with regular and mixed dynamics 
\cite{babl,stru,beta,crli,crpe,tgu,ozha,ssun,hhun} have made it possible 
to describe quantum shell effects in many physical systems in terms of 
the shortest classical periodic orbits (see \cite{book,dpgg} for recent 
surveys). 

However, in all the approaches mentioned so far, the spin degrees of
freedom have not been incorporated in the semiclassical theories. This
becomes, in particular, necessary when one wants to apply the POT to
systems with spin-orbit interactions, such as nuclei, atoms, or 
semiconductor nanostructures. 
Littlejohn and Flynn \cite{lif1} have developed a semiclassical theory 
of systems with multi-component wavefunctions and applied it \cite{lif2} 
to the WKB quantization of integrable spherical systems with the 
standard spin-orbit interaction; Frisk and Guhr \cite{frgu} have
extended their method to deformed systems with spin-orbit interaction.
None of these authors have, however, developed an explicit trace formula.
Bolte and Keppeler \cite{boke} have recently derived a relativistic 
trace formula from the Dirac equation. They studied several non-relativistic
limits and rederived the Littlejohn-Flynn (LF) approach in the limit of a
strong spin-orbit coupling, thereby justifying some {\it ad hoc} 
assumptions made in \cite{frgu}. A problem that has remained unsolved in 
the strong-coupling limit is that of the so-called mode conversion: the 
semiclassical description breaks down in those points (or subspaces) 
of the classical phase space where the spin-orbit interaction locally
becomes zero. 

In the present paper we shall apply the above methods to various two- 
and three-dimensional model systems, test their ability to reproduce the 
coarse-grained quantum-mechanical level densities of these systems, and 
explore their limitations. We shall also discuss the mode conversion 
problem that arises in the strong-coup\-ling limit. Some preliminary 
results of our investigations have been presented in \cite{cham,luso}.

Our paper is organized as follows. After a short reminder of 
semiclassical trace formulae for coarse-grained quantum systems in sec
\ref{sec:trace}, we review in sec \ref{sec:known} the approaches of
\cite{lif1,frgu,boke} which we then apply in the following. In sec 
\ref{sec:2deg0b} we investigate the two-dimensional electron gas (2DEG) 
with a spin-orbit interaction of Rashba type in an external magnetic
field. In sec \ref{sec:2dho} we add to this system a laterally confining 
anisotropic harmonic-oscillator potential as a model for an anisotropic 
semiconductor quantum dot. Section \ref{sec:3dho} is devoted to a 
three-dimensional harmonic oscillator potential with standard spin-orbit 
interaction of Thomas type, as in the shell model for light atomic 
nuclei (see, e.g., \cite{nils}). In sec \ref{sec:mc}, we finally discuss 
the mode conversion problem and present some supporting evidence for the 
diabatic spin-flip hypothesis proposed in \cite{frgu}.

\section{Trace formulae with coarse-graining}
\label{sec:trace}

The primary object of the semiclassical trace formulae is the level 
density (or density of states)
\begin{equation}
g(E) = \sum_k \delta(E-E_k)
\label{gofe}
\end{equation}
of a system described quantum-mechanically by the stationary 
Schr\"odinger equation
\bea
{\widehat H}\, \phi_k = E_k \,\phi_k\,.
\eea
We stay throughout in the non-relativistic limit and assume the 
energy spectrum $\{E_k\}$ to be discrete. Classically, the system is 
described by a Hamilton function $H({\bf r,p})=E$ and the equations of
motion derived from it. The function $H(\bf r,p)$ may be understood 
as the phase-space symbol of the operator ${\widehat H}$ in the limit 
$\hbar\rightarrow 0$.

The level density \eq{gofe} can be written as a sum of a smooth part 
and an oscillating part:
\begin{equation}
g(E)= \widetilde{g}(E)+\delta g(E)\,.
\label{split}
\end{equation}
The smooth part ${\widetilde g}(E)$ is in the POT given by the 
contribution of all orbits with zero length \cite{berm}. It is often 
easily evaluated by the (extended) Thomas-Fermi theory or by a numerical 
Strutinsky averaging \cite{strt} of the quantum spectrum (see \cite{book} 
for the relation of all three methods). The oscillating part $\delta 
g(E)$ is semiclassically approximated by trace formulae of the form
\begin{equation}
\delta g_{sc}(E) = \sum_{po} {\cal A}_{po}(E) 
                   \cos \left(\frac{1}{\hbar} S_{po}(E) 
                   - \frac{\pi}{2} \sigma_{po} \right).
\label{trform}
\end{equation}
The sum is here over all periodic orbits ($po$) of the classical system,
including all repetitions of each primitive periodic orbit ($ppo$).
$S_{po}(E)$ is the action integral and $\sigma_{po}$ the so-called Maslov
index of a periodic orbit. The amplitude ${\cal A}_{po}(E)$ depends on the 
integrability and the continuous symmetries of the system. When all 
periodic orbits are isolated in phase space, the amplitude is given by 
\cite{gutz}
\bea
{\cal A}_{po}(E)=\frac{1}{\pi\hbar}\frac{T_{ppo}} 
                 {\sqrt{|\det (\widetilde{\M}_{po}-\unit)|}},
\label{amppo}
\eea
where $T_{ppo}$ is the period of the primitive orbit and  
$\widetilde{\M}_{po}$ the stability matrix of the periodic orbit.
Examples of amplitude factors for systems with continuous symmetries
or for integrable systems may be found in the literature quoted in the
introduction. 

The $po$ sum in \eq{trform} does not converge in most cases; it must in 
general be understood as a an asymptotic series that is only 
semiconvergent. However, much practical use can be made of trace
formulae if one does not attempt to obtain an exact energy spectrum
(given by the poles of the level density), but if one is interested
only in the {\it coarse-grained} level density. For this, we define a 
smoothed quantum-mechanical level density by a convolution of \eq{gofe}
with a normalized Gaussian:
\begin{equation}
g_\gamma(E)= \frac{1}{\sqrt{\pi}\gamma} 
               \sum_{k} e^{-[(E-E_k)/\gamma]^2}.
\label{gofegam}
\end{equation} 
Here $\gamma$ is a measure of the desired energy resolution. Applying 
the convolution to the right-hand side of \eq{split} will, for small
enough $\gamma$, not affect the smooth part ${\widetilde g}(E)$. The
convolution of the oscillating part, applied to the semiclassical 
trace formula \eq{trform} and evaluating the integration as usual in
the stationary-phase approximation, leads to the the {\it coarse-grained 
trace formula} \cite{stru,sist}
\begin{eqnarray}
\delta g_{sc}(E)= \sum_{po} e^{-(\gamma T_{po}/2\hbar)^2} 
                      {\cal A}_{po}(E)\cos\left(\frac{1}{\hbar}S_{po}(E) 
                      - \frac{\pi}{2} \sigma_{po} \right).
\label{trformgam}
\end{eqnarray}
The only difference to \eq{trform} is the additional exponential factor 
which suppresses the contributions from orbits with longer periods. Due
to this factor, the periodic orbit sum now converges for not too small
values of $\gamma$. Our choice of the Gaussian function in \eq{gofegam}  
is rather arbitrary; cf \cite{boke} where the regularization
of the trace formula is discussed in terms of a general smooth test 
function. In many physical systems, experimentally observable quantum 
oscillations could be well approximated through such coarse-grained 
trace formulae in terms of only a few short periodic orbits (see 
\cite{book,dpgg} for examples). Balian and Bloch \cite{babl} have used
a small imaginary part of the energy, which corresponds to using a
Lorentzian smoothing function.

\section{Periodic orbit theory with spin degrees of freedom}
\label{sec:known}

In our present study, we want to apply the POT to systems of fermions
with spin $s=1/2$, in which the spin degrees of freedom are involved 
through an explicit spin dependence of the Hamiltonian. We write it as
\bea
\widehat{H} =   \widehat{H}_0 \unit + \widehat{H}_1\,, \qquad
\widehat{H}_0 = \frac{\hat{\bf p}^2}{2m} + V({\bf r)}\,,
\label{ham01}
\eea
and assume the spin-dependent part to have the following general form
of a spin-orbit interaction:
\bea 
\widehat{H}_1 = \hbar\,\kappa\,{\widehat{\bf C}}({\bf r},\hat{\bf p})\cdot 
                             {\bgrk{\sigma}}.
\label{hso}
\eea
Here ${\bgrk{\sigma}}=(\sigma_x,\sigma_y,\sigma_z)$ is the vector built
of the three Pauli matrices and $\unit$ is the unit $2\times2$ matrix 
acting in the spin space spanned by the Pauli spinors. The Planck constant 
$\hbar$ in \eq{hso} comes from the spin operator $\hat{\bf s}=\frac12\hbar
\bgrk{\sigma}$. The constant $\kappa$ is such that the spin-orbit term has 
the correct dimension of an energy; it is composed of natural (or material) 
constants but does not contain $\hbar$. The vector ${\bf C}({\bf r},{\bf 
p})$, which is the phase-space symbol of the operator $\widehat{\bf C}
({\bf r},\hat{\bf p})$, may be interpreted as an internal magnetic field 
with arbitrary dependence on the classical phase-space variables ${\bf r},
{\bf p}$. In the standard non-relativistic reduction of the Dirac equation, 
$\widehat{\bf C}({\bf r},\hat{\bf p})$ becomes
\bea
{\widehat{\bf C}}({\bf r},\hat{\bf p})= \left[\,\bgrk{\nabla} V({\bf r}) 
                  \times \hat{\bf p}\,\right],
\label{thomas}
\eea
with $\kappa = 1/4m^2c^2$; $V(\bf r)$ is an external electrostatic 
potential. For the spherical Coulomb potential, \eq{hso} and \eq{thomas}
yield the Thomas term (corrected by a factor 2) well-known in atomic 
physics.

It is by no means trivial, now, how to define a classical Hamiltonian
corresponding to \eq{ham01}, since there is no direct classical analogue 
of the spin. Whereas numerous attempts have been made over the last
seven decades or so to describe the spin classically or semiclassically,
only two approaches have lent themselves to an inclusion of spin degrees
of freedom in semiclassical trace formulae. These are the approaches
developed by Littlejohn and Flynn \cite{lif1,lif2}, with extensions by
Frisk and Guhr \cite{frgu}, and of Bolte and Keppeler \cite{boke}. We
refer to these original papers for all details, as well as for exhaustive
references to the earlier literature. Here we shall briefly present the 
resulting formulae which will be applied and tested for various model 
systems in the following sections.

Bolte and Keppeler \cite{boke} started from the Dirac Hamiltonian to 
derive a relativistic trace formula which, to our knowledge, has not yet
been applied to physical systems. They also started from the 
non-relativistic Pauli equation for a charged particle with spin 1/2 in
an external magnetic field $\bf B(r)$, for which we have
\bea
\widehat{H}_0 = \frac{1}{2m}\left[\hat{\bf p}-\frac{e}{c}{\bf A(r)}\right]^2 
                + V({\bf r})\,,\qquad
\widehat{H}_1 = - \frac{e\hbar}{2mc}\,{\bf B(r)}\cdot\bgrk{\sigma}\,.
\label{hpauli} 
\eea
Using the same techniques as in the derivation of their relativistic trace 
formula, they discuss two limits for introducing the semiclassical 
approximation. The Zeeman term $\widehat{H}_1$ in \eq{hpauli} is not a 
spin-orbit interaction, but Bolte and Keppeler \cite{boke} argue that the 
extension of their methods to the more general form \eq{hso} is straight 
forward. We therefore present their approach below for the general
spin-orbit Hamiltonian \eq{hso}.

\subsection{Weak coupling limit}
\label{sec:wcl}

In the ``weak coupling'' limit (WCL), the semiclassical approximation is
systematically performed by the usual expansion in powers of $\hbar$. 
Because of
the explicit appearance of $\hbar$ in $\H1$, the limit $\hbar\rightarrow 
0$ leads to the classical Hamiltonian $H_{cl}({\bf r,p})=H_0({\bf r,p})$
whose periodic orbits enter the trace formula. The spin degrees of 
freedom here are not coupled to the classical motion. Their contribution 
to the trace formula enters through the trace of a $2\times2$ matrix 
${\dmat}(t)$ obeying the ``spin transport equation''
\begin{eqnarray}
\frac{\rm d}{{\rm d}t}\,{\dmat}(t) 
     = -\i \kappa \left[{\bf C}({\bf r,p}) \cdot \bgrk{\sigma} \right] 
       {\dmat}(t)\,, \qquad {\dmat}(0)=\unit\,,
\label{modfaceq}
\end{eqnarray}
to be evaluated along each periodic orbit ${\bf r}_{po}(t)$, 
${\bf p}_{po}(t)$ found from $H_0({\bf r,p})$. This equation describes 
the spin precession about the instantaneous internal magnetic field 
${\bf C({\bf r,p})}$ along the periodic orbit. Using the solution of 
\eq{modfaceq} for each orbit, the trace formula is given, to leading 
order in $\hbar$, by \cite{boke}
\begin{equation}
\delta g_{sc}(E) = \sum_{po} {\cal A}_{po}(E)\tr {\dmat}(T_{po}) 
                   \cos \left(\frac{1}{\hbar} S_{po}(E) 
                   - \frac{\pi}{2} \sigma_{po} \right),
\label{trwcl}
\end{equation}
where $T_{po}=$ d$S_{po}(E)$/d$E$ is the period of each (repeated) orbit.
Since the periodic orbits are not affected by the spin motion, the only
difference to the standard trace formula \eq{trform} is the appearance
of the spin modulation factor $\tr {\dmat}(T_{po})$; all other 
ingredients are evaluated in the usual manner for the unperturbed 
periodic orbits of $H_0$. One may therefore consider this treatment as 
an adiabatic limit of fast spin motion and slow spatial motion 
${\bf r}(t)$, but Bolte and Keppeler \cite{boke} argue that this 
adiabatic assumption is not needed for the formula \eq{trwcl} to be 
true to leading order in $\hbar$. 

Through the factor $\tr{\dmat}(T_{po})$ the contribution of a given 
periodic orbit depends on the overlap of the spin directions at the 
beginning and the end of its period. An orbit for which these two 
directions are identical has simply $\tr{\dmat}(T_{po})=2$, whereas 
an orbit for which these directions are antiparallel does not contribute 
at all to the trace formula \eq{trwcl}. 

For self-retracing orbits, i.e., librations between two turning points 
in coordinate space, the spin precession described by \eq{modfaceq} is 
reversed at each turning point, and hence the spin direction is brought 
back to its initial value after a full period. Such orbits therefore 
only acquire a trivial factor $\tr{\dmat}(T_{po})=2$ compared to the 
trace formula \eq{trform}. For systems which possess only 
self-retracing periodic orbits (see, e.g., the examples in secs 
\ref{sec:3dho} and \ref{sec:mc}), the formula \eq{trwcl} thus reduces 
to the trivial recipe of incorporating the spin by a simple degeneracy 
factor 2 in the level density, which cannot account for the spin-orbit
interaction effects.

\subsection{Strong coupling limit}
\label{sec:scl}

To obtain the ``strong coupling'' limit (SCL), Bolte and Keppeler 
\cite{boke} follow the philosophy of \cite{lif2,frgu} by absorbing 
the Planck constant $\hbar$ in \eq{hpauli} into the Bohr magneton 
$\mu = e\hbar/2mc$, thus considering $\mu$ as a constant in the 
semiclassical limit $\hbar\rightarrow 0$. Similarly, for the spin-orbit
Hamiltonian \eq{hso} one absorbs $\hbar$ into the constant 
${\bar\kappa}=\hbar\kappa$. The fact that this corresponds to a double 
limit $\hbar\rightarrow 0$ and $\kappa\rightarrow\infty$, with ${\bar 
\kappa}=\hbar\kappa$ kept constant, justifies the name ``strong 
coupling'' limit.

The symbol of the full Hamiltonian in phase space now remains a 
$2\times2$ matrix which after diagonalization leads to the two 
classical Hamiltonians
\bea
H_\pm({\bf r,p})=H_0({\bf r,p}) \pm {\bar\kappa}\,|{\bf C}({\bf r,p})|\,,
\label{hpm}
\eea
which can be considered as two adiabatic Hamiltonians with opposite spin 
polarizations. They create two classes of dynamics, whose periodic orbits 
must be superposed in the final trace formula. Such a trace formula has, 
however, not been derived explicitly so far. Littlejohn and Flynn 
\cite{lif1} argued that a non-canonical transformation of the phase-space 
variables $\bf r,p$ would be necessary to calculate the amplitudes. Frisk 
and Guhr \cite{frgu} surmised, based upon Fourier transforms of quantum 
spectra, that this is not necessary, provided that the actions $S^\pm_{po}$ 
of the periodic orbits generated by the Hamiltonians $H_\pm$ be corrected 
by some phases accumulated along the periodic orbits:
\begin{eqnarray}
\frac{1}{\hbar}\,S^\pm_{po} \rightarrow
                  \frac{1}{\hbar}\,S^\pm_{po}+\Delta\Phi_\pm\,, \qquad
\Delta\Phi_\pm = \oint_{po} (\lambda^{\mathrm{B}}_\pm 
                                 + \lambda^{\mathrm{NN}}_\pm) \,\d t\,.
\label{deltaS}
\end{eqnarray}
The phase velocities $\lambda^{\mathrm{B}}_\pm,\;\lambda^{\mathrm{NN}}_\pm$,
which have been called the ``Berry'' and the ``no-name'' terms 
\cite{lif1,lif2}, arise as first-order $\hbar$ corrections in the 
semiclassical expansion of the symbol of the Hamiltonian matrix. 
Bolte and Keppeler \cite{boke} have used their techniques to give this 
prescription a rigorous justification. For the Hamiltonians \eq{hpm} with 
\eq{thomas}, the above phase velocities can be calculated most easily in 
terms of the polar angles $\theta,\phi$ defining the unit vector of the 
instantaneous direction of $\bf C({\bf r,p})$, i.e., ${\bf e}_C=
(\cos\phi\sin\theta,\sin\phi\sin\theta,\cos\theta)$, and of the Hesse 
matrix of the potential, {\large \tt V}$''_{ij}=\partial^2 
V({\bf r})/\partial r_i\partial r_j$ ($i,j=x,y,z$), evaluated along the 
periodic orbits, and are given by 
\cite{frgu} 
\begin{eqnarray}
\lambda_\pm^{\mathrm{B}}=\mp\,\frac{1}{2}\,(1-\cos\theta)\,\dot{\phi}\,,
                          \qquad
\lambda_\pm^{\mathrm{NN}}=-\frac{\bar\kappa}{2}\,{\bf e}_C^T\,
                           \hbox{\large\tt V}''{\bf e}_C\,.
\label{phases}
\end{eqnarray}

Clearly, in the SCL the spin affects the classical dynamics, albeit only 
in an adiabatic, polarized fashion. Moreover, there is a serious
limitation to the procedure outlined above. Whenever $\bf C=0$ at a given 
point in (or in a subspace of) phase space, the two Hamiltonians $H_\pm$ 
become degenerate and singularities arise, both in the classical 
equations of motion and in the calculation of the phase corrections 
\eq{phases} and the stabilities of the periodic orbits. Such points are 
called ``mode conversion'' (MC) points. A similar situation occurs in the 
chemistry of molecular reactions when two or more adiabatic surfaces 
intersect. The MC poses a difficult problem in semiclassical physics and 
chemistry, that has not been satisfactorily solved so far for systems 
with more than one spatial dimension (see \cite{mc1d} for a discussion of 
the MC in one dimension).

For self-retracing periodic orbits, all components of the momentum $\bf p$
are zero at the turning points. Hence, a spin-orbit interaction of the
standard Thomas type \eq{thomas} will in the SCL lead to MC at the
turning points. In this case, both the WCL and the SCL break down, and
an improved treatment becomes necessary to include such orbits in a
semiclassical trace formula. A new approach that is free of the MC problem
has just been proposed \cite{plet}; its results will be presented in a 
forthcoming paper. 

\newpage

\section{Two-dimensional electron systems with Rashba term}
\label{sec:2deg}

In this section we shall investigate a two-dimensional electron gas 
(2DEG) with a spin-orbit interaction of the Rashba type \cite{bych}.
We will include also an external magnetic field and an external potential
$V({\bf r})=V(r_x,r_y)$ which causes a lateral confinement of the 2DEG 
(see, e.g., \cite{darn}). The Rashba term can be written in the form 
\begin{eqnarray}
{\widehat H}_1=\hbar\kappa\,\widehat{\bf C}({\bf r},\hat{\bf p})\cdot
                            \hat{\bgrk{\sigma}},\qquad 
\widehat{\bf C}({\bf r},\hat{\bf p}) = \left( \begin{array}{c}
                             -\langle v_z' \rangle\, \hat{p}_y\\ 
                            ~~\langle v_z' \rangle\, \hat{p}_x\\
                  \hat{p}_y\,\partial V({\bf r})/\partial r_x -
                  \hat{p}_x\,\partial V({\bf r})/\partial r_y\! 
                  \end{array}\right). 
\label{rashba} 
\end{eqnarray}
Here $\langle v_z'\rangle$ is the mean gradient of the electrostatic
potential in the $z$ direction that confines the electron gas to the 
$(x,y)$ plane, so that $z=p_z=0$, and the constant $\kappa$ depends on
the band structure of the semiconductor in which the 2DEG is confined
\cite{darn}. Note that the Rashba term \eq{rashba} is of the standard 
form \eq{thomas}. 

We shall in sec \ref{sec:2deg0b} study the case of the free 2DEG for 
$V({\bf r})=0$ in an external perpendicular homogeneous magnetic field,
for which a quantum-mechanically exact trace formula is easily derived. 
We shall see that the WCL approach yields an analytical semiclassical trace 
formula which is exact only to leading order in $\hbar\kappa$ and in the 
limit $\kappa\rightarrow 0$. The SCL approach, however, for which we also 
obtain an analytical result, is demonstrated to fail for $\kappa\rightarrow 
0$, but to include correctly the higher-order terms in $\hbar\kappa$, and 
to become exact in the strong-coupling limit. In sec \ref{sec:2dho}, we 
add an anisotropic harmonic confinement potential $V(\bf r)$ to this 
system, where the WCL can be applied successfully in numerical 
calculations. When the external magnetic field is switched off, the 
remaining system has only self-retracing orbits for which both the WCL 
and the SCL fail. The MC problem arising in the SCL for this system 
will be discussed in section \ref{sec:mc}.

\subsection{Free 2DEG with Rashba term in an external magnetic field}
\label{sec:2deg0b}

We first discuss the free 2DEG with the Rashba spin-orbit interaction 
\eq{rashba} in a homogeneous magnetic field ${\bf B}=B_0\,{\bf e}_z$. 
The corresponding vector potential $\bf A$ is included by replacing 
the momentum operator in the usual way: $\hat{\bf p} \rightarrow 
\hat{\bgrk{\pi}} = \hat{\bf p} - e {\bf A}/c$. The total Hamiltonian 
then reads
\bea
\hat{H} = \frac{\hat{\bgrk{\pi}}^2}{2m^*}\unit+\hbar\kappa\,
        \widehat{\bf C}({\bf r},\hat{\bgrk{\pi}})\cdot{\bgrk{\sigma}},
        \qquad\qquad
\widehat{\bf C}({\bf r},\hat{\bgrk{\pi}}) 
                 = \left( \begin{array}{c} \!\!-\langle v_z' 
                   \rangle\,\hat{\pi}_y \\ \!\!\langle v_z' 
                   \rangle\,\hat{\pi}_x\!\!\! \\ 0 \end{array}\!\right),
\label{free2deg}
\eea 
where, using the symmetric gauge for ${\bf A}$, the mechanical momentum
is given by 
\bea
\hat{\bgrk{\pi}} = \left( \begin{array}{c} \hat{p}_x \\\hat{p}_y \\ 0 
                   \end{array} \!\right) - \frac{eB_0}{2c} 
                   \left(\!\! \begin{array}{c} -r_y \\ r_x\!\! \\ \,0 
                   \end{array} \!\!\right).
\eea
Here $m^*$ is the effective mass of the electron and $e$ its charge. 
We have omitted the spin contribution to the Zeeman term which could be 
trivially included by adding the magnetic field to $\widehat{\bf C}$.

The quantum-mechanical eigenvalues of the Hamiltonian \eq{free2deg} 
are known analytically \cite{bych}:
\begin{eqnarray}
E_0 = \hbar\omega_c/2\,, \qquad \qquad
E^\pm_n = \hbar\omega_c\left(n\pm\sqrt{1/4
          +\hbar\,2n\wkap^2}\,\right), \quad n=1,2,3,\dots 
\end{eqnarray}
Here $\omega_c=eB_0/m^*\!c$ is the cyclotron frequency, and 
$\wkap=\kappa\,\langle v'_z\rangle\sqrt{m^*\!/\omega_c}$ is a 
renormalized coupling constant which we have defined in such a way 
that the Planck constant $\hbar$ appears explicitly in all our 
formulae.\footnote{In the literature on the Rashba term, the constant 
$\alpha=\hbar^2\kappa\langle v'_z\rangle$ is frequently used, see 
\cite{bych,darn}} The exact level density is then given by
\bea
g(E)=\delta(E-E_0)+\sum_{n=1}^\infty\,[\,\delta(E-E_n^+)
                                       +\delta(E-E_n^-)\,]\,.
\label{grash}
\eea
Using Poisson summation (see, e.g., \cite{book}, sec 3.2.2), this result 
can be identically transformed to an exact quantum-mechanical trace formula. 
The smooth part of \eq{grash} is ${\widetilde g}(E)=2/\hbar\omega_c$, 
and the oscillating part becomes
\begin{eqnarray}
\delta g(E) & = & \frac{2}{\hbar\omega_c}\! \sum_\pm
             \biggl(1\pm\frac{\hbar\wkap^2}{\sqrt{1/4+2E\,\wkap^2\!/\omega_c
                                        +\hbar^2\wkap^4}}\biggr)\nonumber\\
            &   &  \qquad \quad \times \sum_{k=1}^\infty
             \cos\!\left[ k\,2\pi\! \left( \frac{E}{\hbar\omega_c} 
             +\hbar\,\wkap^2\pm\sqrt{1/4+2E\,\wkap^2\!/\omega_c+\hbar^2\wkap^4}
             \,\right)\!\right].
\label{tracefromqm}
\end{eqnarray}

We will now analyze the system semiclassically, using both the WCL and the 
SCL approach. In the weak-coupling limit, we first need the trace formula 
for the unperturbed system without spin-orbit coupling, corresponding to 
${\widehat H}_0=\hat{\bgrk{\pi}}^2\!/2m^* $. This is the quantized Landau 
level system, whose exact trace formula is that of a one-dimensional
harmonical oscillator with the cyclotron frequency $\omega_c$ and reads 
\cite{book} (without spin degeneracy factor)
\begin{eqnarray}
\delta g^{(\kappa=0)}(E)= \frac{2}{\hbar \omega_c}\sum_{k=1}^{\infty} (-1)^k
\cos \left( k \frac{2\pi E}{\hbar\omega_c}\,\right).
\label{trace1ho}
\end{eqnarray}
For the Rashba term in \eq{free2deg}, the spin transport equation 
\eq{modfaceq} can be solved analytically \cite{cham}, and the spin 
modulation factor becomes
\begin{eqnarray}
\tr{\dmat}(k T_{po})=(-1)^k\,2\cos\!\left[ k\,2\pi\sqrt{1/4
                     +2E\,\wkap^2\!/\omega_c}\,\right].
\end{eqnarray}
With \eq{trwcl}, the complete semiclassical trace formula in the WCL
can therefore be written as
\begin{eqnarray}
\delta g_{sc}^{WCL}(E)=\frac{2}{\hbar\omega_c}\!\sum_\pm \sum_{k=1}^\infty
                 \cos\!\left[ k\,2\pi\! \left( \frac{E}{\hbar\omega_c} 
                 \pm\sqrt{1/4+2E\,\wkap^2\!/\omega_c}\,\right)\!\right].
\label{tracewcl}
\end{eqnarray} 
This result is not the same as the exact quantum-mechanical one 
\eq{tracefromqm}, but it contains the correct terms of leading 
order in $\hbar$, in accordance with the derivation of Bolte and 
Keppeler \cite{boke}, and the correct leading-order term in $\wkap^2$. 
The missing terms would come about by going to higher orders in the 
semiclassical $\hbar$ expansion. Note also that \eq{tracewcl} 
becomes exact in the limit $\wkap\rightarrow0$.

Recently, Keppeler and Winkler \cite{wink} have analyzed the anomalous
magnetoresistance oscillations of a quasi-2DEG in GaAs semiconductors,
employing two kinds of spin-orbit interactions one of which was of the
Rashba type \eq{free2deg}. They applied the WCL trace formula 
\eq{trwcl} and obtained good agreement with quantum-mechanical results.
As the spin-orbit interaction in GaAs is rather weak, we assume that
their results were not sensitive to the missing higher-order 
semiclassical terms, which explains their good agreement.  

It is very instructive now to compare the above result with that of an 
analysis using the strong-coupling limit. The SCL Hamiltonians \eq{hpm} 
become $H_\pm({\bf r,p})=H_0({\bf r,p})\pm{\bar\wkap}
\sqrt{2\omega_cH_0({\bf r,p})}$ with $\bar\wkap=\hbar\wkap$. It is easy 
to see that $H_0({\bf r,p})=E_0$ is a constant of motion. The equations 
of motion derived from $H_\pm$ therefore become linear, representing 
one-dimensional harmonic oscillators like for the Landau orbits of the
unperturbed system $H_0$. Instead of the cyclotron frequencies $\omega_c$
they have, however, the modified eigenfrequencies
\bea
\omega_\pm = \omega_c\left(1\pm\bar\wkap\sqrt{\omega_c/2E_0}\right).
\label{ompm}
\eea
Since ${\bf C}({\bf r,p})$ does not change its sign along the modified 
Landau orbits, the system does not suffer from the mode conversion
problem and the SCL can be safely applied. The action integrals of the
primitive orbits are simply found to be $S_\pm=2\pi E_0/\omega_c$, like 
for the unperturbed system, but expressing them in terms of the conserved 
total energy $E=E_0\pm{\bar\wkap}\sqrt{2\omega_cE_0}$ one finds
\bea
S_\pm(E) = 2\pi\!\left(\frac{E}{\omega_c}+{\bar\wkap}^2
       \mp \sqrt{2E\,{\bar\wkap}^2\!/\omega_c+{\bar\wkap}^4}\,\right).
\eea 
The phase velocities \eq{phases} are easily found to be $\lambda_\pm^B
=\mp\,{\dot\phi}/2=\pm\,\omega_\pm/2$ and $\lambda_\pm^{NN}=0$, so that 
the overall phase correction \eq{deltaS} becomes $\Delta\Phi_\pm=\pm\,\pi$ 
for each repetition of the primitive orbits. Inserting these results into 
the trace formula of the one-dimensional harmonic oscillator, i.e., eq
\eq{trace1ho} with $\omega_c$ replaced by $\omega_\pm$, using
$T_\pm=2\pi/\omega_\pm=\d S_\pm/\d E$, and summing over both orbit 
types, we obtain the semiclassical trace formula in the SCL (exhibiting
again the $\hbar$ contained in $\bar\wkap$)
\begin{eqnarray}
\delta g_{sc}^{SCL}(E) & = & \frac{2}{\hbar\omega_c}\! \sum_\pm
             \biggl(1\pm\frac{\hbar\wkap^2}{\sqrt{2E\,\wkap^2\!/\omega_c
             +\hbar^2\wkap^4}}\biggr)\nonumber\\
            &   &  \qquad \quad \times \sum_{k=1}^\infty
             \cos\!\left[ k\,2\pi\! \left( \frac{E}{\hbar\omega_c} 
             +\hbar\,\wkap^2\pm\sqrt{2E\,\wkap^2\!/\omega_c+\hbar^2\wkap^4}
             \,\right)\!\right].
\label{tracescl}
\end{eqnarray}
It is interesting to note that hereby the Berry term, yielding the phase
correction $k\Delta\Phi_\pm=\pm\,k\pi$, cancels the alternating sign 
$(-1)^k$ in \eq{trace1ho}. We note that the result \eq{tracescl} would 
be exactly identical to the quantum-mechanical result \eq{tracefromqm}, 
were it not for the missing term 1/4 under the roots. We see, therefore, 
that the SCL result will fail in the limit $\wkap\rightarrow 0$, since 
the alternating sign $(-1)^k$ arises precisely from that missing term 
1/4 in the actions. On the other hand, the SCL trace formula 
\eq{tracescl} does correctly include the higher-order terms in 
$\hbar\wkap^2$, both in actions and amplitudes, becoming exact in the 
limit of a large spin-orbit coupling parameter $\wkap$, as could be 
hoped. Note that \eq{tracescl} 
becomes exact also in the limit of large energy $E$. That the SCL trace 
formula fails in the limit $\wkap\rightarrow 0$ is not surprising 
because of the non-analytic behaviour of the Hamiltonians \eq{hpm}, as 
already pointed out in \cite{lif1}.

\subsection{A quantum dot with external magnetic field}
\label{sec:2dho}

We now add a lateral confining potential $V(r_x,r_y)$ to the previous
system. This is a simple model for a two-dimensional quantum dot which
nowadays can easily be manufactured in semiconductor heterostructures.
We choose the confining potential to be harmonic, so that the Hamiltonian
becomes
\begin{eqnarray}
\hat{H} = \frac{\hat{\bgrk{\pi}}^2}{2m^*}\unit + \frac{m^*}{2}\!\left(
          \omega_x^2\,r_x^2+\omega^2_y\,r_y^2\right)\!\unit 
        + \hbar\kappa\,\widehat{\bf C}({\bf r},\hat{\bgrk{\pi}})\cdot 
          \hat{\bgrk{\sigma}}.
\label{osc2deg}
\end{eqnarray}
The Rashba term $\widehat{\bf C}$ now acquires also a $z$ component like
in \eq{rashba} and reads:
\begin{eqnarray}
\widehat{\bf C}({\bf r},\hat{\bgrk{\pi}}) = \left( 
\begin{array}{c} -\langle v_z' \rangle \hat{\pi}_y\\ \langle v_z' \rangle 
\hat{\pi}_x \\ m^* \omega_x^2r_x \hat{\pi}_y - m^* \omega_y^2 r_y
\hat{\pi}_x
\end{array} \right).
\end{eqnarray}
In the case where the two oscillator frequencies $\omega_x$ and $\omega_y$ 
are identical, the total system has axial symmetry and is integrable even 
including the spin-orbit term, with the eigenvalues of the total angular 
momentum in $z$ direction, ${\hat J}_z={\hat L}_z+{\hat s}_z$, being 
constants of the motion. An exact trace 
formula can then be found by the EBK quantization following the methods 
of \cite{lif2}. We will not discuss the integrable system here and refer 
the interested reader to \cite{cham}. A less trivial situation arises when 
the frequencies $\omega_x$ and $\omega_y$ are different and the system 
with spin-orbit coupling is no longer integrable. 

Since a realistic spin-orbit coupling in most semiconductors is weak, wee 
shall here only use the WCL to derive a semiclassical trace formula for the 
Hamiltonian \eq{osc2deg}. The system without spin-orbit coupling is
biquadratic in the space and momentum variables and can be transformed 
to become separable in its normal modes. The normal-mode frequencies are 
\bea
\omega_\pm = \left[\,\frac{1}{2}\!\left(\omega_c^2+\omega_x^2+\omega_y^2 
             \pm \sqrt{(\omega_c^2+\omega_x^2+\omega_y^2)^2
             -4\,\omega_x^2\omega_y^2}\,\right) \right]^{1/2}.
\eea
The exact eigenenergies are thus given in terms of two oscillator
quantum numbers $n_+$ and $n_-$:
\bea
E_{n_+,n_-} = \hbar \omega_+\! \left( n_+ + 1/2 \right)
             +\hbar \omega_-\! \left( n_- + 1/2 \right) 
              \qquad n_\pm=0,1,2,\dots
\eea
The semiclassical trace formula for such a system is known \cite{bj95}
and quantum-mechanically exact:
\begin{eqnarray}
\delta g^{(\kappa=0)}(E)&=& \sum_\pm \,\frac{1}{\hbar\omega_\pm}
                            \sum_{k=1}^\infty \,(-1)^k
              \frac{1}{\sin\left[k\pi(\omega_\mp/\omega_\pm)\right]} 
              \,\sin\left(\!k\,\frac{2\pi E}{\hbar\omega_\pm}\right).
\label{osctrform}
\end{eqnarray}
Note that this formula is only useful when $\omega_+/\omega_-$ is 
irrational. For rational frequency ratios, careful limits must be 
taken to cancel all singularities, see \cite{bj95} for details. However,
for a finite external field $B_0\ne0$, the ratio $\omega_+/\omega_-$
can always be made irrational by an infinitesimal change of the field
strength, so that equation \eq{osctrform} is adequate for all practical
purposes. The semiclassical origin of this trace formula is given by 
the existence of only two isolated rotating orbits with frequencies
$\omega_+$ and $\omega_-$, whose shapes in coordinate space are
ellipses. Each orbit contributes one of the above two sums; $k$ is the 
repetition number of the primitive orbits (which have $k=1$).

We next have to calculate the spin modulation factors by solving the 
equation \eq{modfaceq} along the unperturbed elliptic orbits. This 
could only be done numerically. It is, however, sufficient to 
calculate the modulation factors for the primitive orbits only. 
Using the property $\tr {\dmat}(k T_{po})=\tr {\dmat}^k(T_{po})$, 
the final trace formula in the WCL is then given by:
\begin{eqnarray}
\delta g_{sc}(E)&=& \sum_\pm \,\frac{1}{\hbar\omega_\pm}
                    \sum_{k=1}^\infty \,(-1)^k 
              \frac{\tr {\dmat}^k(T_\pm)}
              {\sin\left[k\pi(\omega_\mp/\omega_\pm)\right]} 
              \,\sin\left(\!k\,\frac{2\pi E}{\hbar\omega_\pm}\right),
\label{osctrform2}
\end{eqnarray}
where $T_\pm=2\pi/\omega_\pm$ are the periods of the unperturbed
primitive orbits.

In fig \ref{2degoscfig} we compare results for the oscillating parts 
$\delta g(E)$ of the coarse-grained level density for 
$\gamma=0.3\,\hbar\omega_0$; all energies are in units of 
$\hbar\omega_0$. The deformation of the confinement potential was 
fixed by $\omega_x=\omega_0$ and $\omega_y=1.23\,\omega_0$, and the 
cyclotron frequency was chosen to be $\omega_c=0.2\,\omega_0$. In our 
numerical calculations we have set $\hbar=\omega_0=m^*=e=c=\langle 
v_z'\rangle=1$. In these units, the spin-orbit coupling parameter was 
chosen to be $\kappa=0.1$. The heavy solid lines in the upper three 
panels represent the full quantum-mechanical result obtained from an 
exact diagonalization of the Hamiltonian \eq{osc2deg} in the basis
of ${\widetilde H}_0$. In the top panel, the semiclassical trace formula 
\eq{osctrform} without spin-orbit interaction is shown (only $k=1$ and 
$k=2$ contribute visibly). It clearly demonstrates that the effect of 
the spin-orbit interaction on the level density, even at this resolution,
is dramatic. In the next two panels, the spin-orbit interaction has been 
included into the semiclassical WCL trace formula \eq{osctrform2}, using 
the numerical spin modulation factors. We see that the agreement is 
improved radically, especially if the second repetitions ($k=2$) are added. 
The difference between quantum mechanics (QM) and semiclassics (SC) can
clearly be seen only in the close-up (second panel), which selects the
energy region $11 \siml E/(\hbar\omega_0) \siml 21$, where the 
disagreement actually is worst. The bottom panel shows the energy 
dependence of the two spin modulation factors $\tr{\dmat}(T_+)$ and 
$\tr{\dmat}(T_-)$ of the primitive orbits. Clearly, the strong
long-range modulation in the amplitude of $\delta g(E)$ is the result of 
the spin-orbit interaction; it is correctly reproduced in the WCL
approach through the inclusion of the spin modulation factors.

\Figurebb{2degoscfig}{0}{0}{993}{1105}{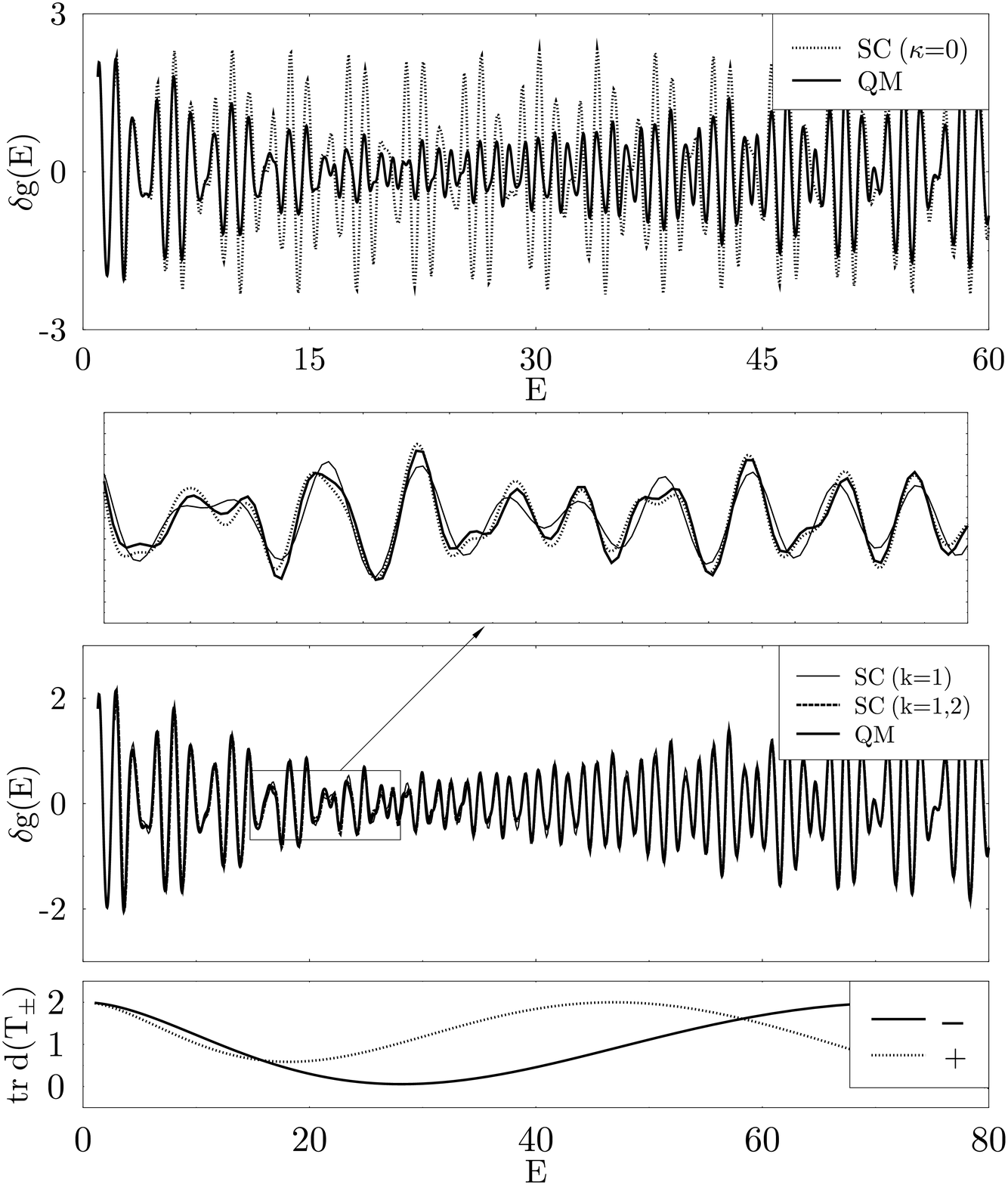}{12.5}{15}{
{\it Upper three panels:} Coarse-grained (with $\gamma=0.3\,\hbar\omega_0$)
oscillating part $\delta g(E)$ of level density of the 2-dimensional 
quantum dot with Rashba term (energy units: $\hbar\omega_0$). Heavy solid
lines: full quantum results (QM) including Rashba term. Dotted line in the
top panel: trace formula \eq{osctrform} for $\kappa=0$. Solid and dashed
lines in 2nd and 3rd panel: semiclassical trace formula \eq{osctrform2}
(SC) with only first ($k=1$) and up to second harmonics $(k=2)$ included.
{\it Lowest panel:} spin modulation factors $\tr{\dmat}(T_+)$ and
$\tr{\dmat}(T_-)$.
}

\vspace*{-0.5cm}

This concludes the discussion of the system with magnetic field $B_0\neq0$.
We notice at this point that the case $B_0=0$ with irrational frequency 
ratio $\omega_x/\omega_y$ is not accessible in the WCL. This follows from 
the fact that in the system without spin-orbit coupling, the only periodic 
orbits are the self-retracing librations along the principal axes. As 
already discussed at the end of sec \ref{sec:wcl}, the WCL fails here 
in that it gives only the trivial modulation factor tr$\,{\dmat}(T_{po})=2$ 
for both orbits. On the other hand, the SCL suffers from the MC problem.
We will discuss this problem explicitly in sec \ref{sec:mc}, where we
return to the above system with $B_0=0$.


\section{Three-dimensional harmonic oscillator with standard spin-orbit
         interaction}
\label{sec:3dho}

We now discuss a three-dimensional system with a spin-orbit interaction 
of the Thomas type, as it is well-known from non-relativistic atomic and 
nuclear physics. In order to be able to perform as many calculations as
possible analytically, we choose again a harmonic-oscillator potential
for $V({\bf r})$. This potential is not only the prototype for any
system oscillating harmonically around its ground state, but it has
actually been used in nuclear physics as a realistic shell 
model\footnote{The angular-momentum dependent ${\hat L}^2$ term included 
in the Nilsson model \cite{nils} is of minor importance in light nuclei; 
it is left out here to simplify our investigation} for light nuclei 
\cite{nils}, provided that the spin-orbit interaction was included with 
the correct sign \cite{gmay}.

We thus start from the following Hamiltonian
\bea
\widehat{H} = \widehat{H}_0 \unit
            + \hbar\kappa\, \hat{\bf C}({\bf r},\hat{\bf p}) 
              \cdot \hat{\bgrk{\sigma}}\,,\qquad
 \hat{\bf C}({\bf r},\hat{\bf p}) = \nabla V({\bf r})\times\hat{\bf p}\,, 
\label{sysgen}
\eea
with
\bea
\widehat{H}_0 = \frac{1}{2}\,\hat{\bf p}^2 + V({\bf r})\,, \qquad\quad 
 V({\bf r}) = \sum_{i=x,y,z} \frac12\,\omega_i^2 r_i^2\,.
\label{3dh0}
\eea
Here ${\bf r}=(r_x,r_y,r_z)$ and ${\bf p}=(p_x,p_y,p_z)$ are 
three-dimensional vectors. We express the three oscillator frequencies
in terms of two deformation parameters $\alpha$, $\beta$:
\begin{eqnarray}
\omega_x=\omega_0\,,\qquad \omega_y=(1+\alpha)\,\omega_0\,,\qquad
\omega_z=(1+\alpha)^\beta \omega_0\,.
\label{defpar}
\end{eqnarray}
and use $\hbar\omega_0$ as energy units. For $\alpha=0$, $\beta=1$ the 
system has spherical symmetry, for $\beta=1$ and $\alpha\ne0$ it has
only axial symmetry. $\kappa$ will be measured in units of 
$(\hbar\omega_0)^{-1}$.

We shall first (sec \ref{ssec:3dqm}) briefly discuss the quantum-mechanical 
spectrum of the system, and then (sec \ref{ssec:3dsc}) investigate it in
more detail by the semiclassical methods. The most interesting case is
that where the oscillator frequencies are mutually irrational, so that
the unperturbed classical Hamiltonian $H_0$ has only self-retracing
periodic orbits. In this case the WCL cannot handle the spin-orbit
coupling, and we must resort to the SCL. As we will show, the leading
periodic orbits with shortest periods in this case do not undergo mode
conversion. We therefore use this system for a representative case study, 
for which a trace formula can be successfully derived (cf \cite{luso,cham}) 
within the SCL.

\subsection{Quantum-mechanical spectrum}
\label{ssec:3dqm}
 
In general, the system \eq{sysgen} is not integrable. There are, however,
two well-known integrable cases for which the quantum spectrum is 
analytically known: the separable system \eq{3dh0} without coupling 
($\kappa=0$), and the spherical system ($\alpha=0$) including coupling.
The unperturbed harmonic oscillator has the spectrum
\begin{eqnarray}
E_{n_xn_yn_z}= \sum_{i=x,y,z} \hbar \omega_i \,(n_i + 1/2 \,)\,.
\qquad n_i=0,1,2,\dots
\label{lev3dh0}
\end{eqnarray}
The spherical system ($\alpha=0$) with coupling $\kappa\neq0$, for which
$\hat{\bf C}({\bf r},\hat{\bf p})=\omega_0^2\,\hat{\bf L}$, has the 
spectrum
\begin{eqnarray}
E_{nlj}= \hbar\omega_0\,(2n+l+3/2) +\kappa\,(\hbar\omega_0)^2\times 
\left\{\!\! \begin{array}{cc} l & \; \mbox{for} \quad j=l+1/2\,,\\ 
                             -(l+1) & \; \mbox{for} \quad j=l-1/2\,, 
            \end{array} \right.
\label{specso0}
\end{eqnarray}
where $n,l=0,1,2,\dots$ and $j=l\pm 1/2$ is the total angular momentum. 
The spin-orbit term is only to be included for $l>0$. Each level
$E_{nlj}$ has the usual angular momentum degeneracy $(2j+1)$ which
equals 2 for $l=0$. Note that in nuclear physics, $\kappa$ is negative
\cite{nils,gmay}.

The non-integrable cases require numerical methods for determining
the energy spectrum. Here we used the diagonalization in the basis 
$\i^{n_y}|n_x,n_y,n_z, s_z\rangle$ of the unperturbed Hamiltonian 
$\widehat{H}_0$ \eq{3dh0} with the eigenenergies \eq{lev3dh0}, where 
$|s_z\rangle$ with $s_z=\pm 1$ are the spin eigenstates of 
$\hat{\sigma}_z$. The inclusion of the phase $\i^{n_y}$ leads to real 
matrix elements; furthermore, the conservation of the signatures 
$(-1)^{n_x+n_y+n_z}$ and $(-1)^{n_x+n_y}s_z$ allows one to separate 
the Hamiltonian matrix into smaller uncoup\-led blocks (see \cite{gotz}
for details). 

\subsection{Semiclassical analysis}
\label{ssec:3dsc}

\subsubsection{Smooth level density}

When one wants to compare results of semiclassical trace formulae
with quantum-mechanical level densities, one has to subtract from
the latter the smooth part $\widetilde{g}(E)$ (see sec 
\ref{sec:trace}). For the Hamiltonian \eq{sysgen}, $\widetilde{g}(E)$ 
can be calculated analytically within the extended Thomas-Fermi (TF) 
method, which has been done already long ago \cite{je75}. The result, 
as an expansion both in $\hbar$ and powers of $\kappa$, reads
\begin{eqnarray}
\widetilde{g}(E)\!\!& = &\! \frac{E^2}{\hbar^3\,\omega_x\omega_y\omega_z}
                      \Bigl\{1
                      +\hbar^2\kappa^2\,(\omega_x^2+\omega_y^2+\omega_z^2)
                      +{\cal O}(\hbar^4\kappa^4)\Bigr\}\nonumber\\
                      \nonumber\\ 
                & + &\! \frac{2E}{3\,\hbar^2\,\omega_x\omega_y\omega_z}
                      \Bigl\{\hbar^3\kappa^3\,(\omega_x^2\omega_y^2
                      +\omega_y^2\omega_z^2+\omega_z^2\omega_x^2)
                      +{\cal O}(\hbar^5\kappa^5)\Bigr\}\nonumber\\
                      \nonumber\\ 
                & - &\! \frac{(\omega_x^2+\omega_y^2+\omega_z^2)}
                           {12\,\hbar\,\omega_x\omega_y\omega_z}
                      \biggl\{1+\hbar^2\kappa^2\,
                      \frac{(\omega_x^2+\omega_y^2+\omega_z^2)^2
                            +2\,(\omega_x^2\omega_y^2+\omega_y^2
                            \omega_z^2+\omega_z^2\omega_x^2)}
                           {(\omega_x^2+\omega_y^2+\omega_z^2)}
                      +{\cal O}(\hbar^4\kappa^4)\biggl\}.~~~~~~
\label{smooth}
\end{eqnarray}
In the literature, the smooth part is often assumed to be given 
by the TF model. This leads, however, only to the leading term 
proportional to $E^2$. For an accurate determination of 
$\widetilde{g}(E)$, the leading $\hbar$ and $\hbar^2$ corrections 
relative to the TF term may not be neglected.

\subsubsection{Trace formulae for the integrable cases}

The exact spectra of the integrable cases offer again the possibility 
to derive trace formulae that are exact in all orders of $\hbar$. For
the unperturbed harmonic oscillators, these are known \cite{book,bj95}
and need not be repeated here. For the spherical case with spin-orbit
interaction, the methods of \cite{book,bj95,je75} lead to the following
result: 
\begin{eqnarray}
\delta g(E) & = & \frac{E}{(\hbar\omega_0)^2}\sum_{\pm}\sum_{k=1}^{\infty} 
                  \frac{1}{(1\pm\kappa\hbar\omega_0)^2}\frac{1}
                  {\sin\left[2k\pi/(1\pm\kappa\hbar\omega_0)\right]}
                  \,\sin\left(k\,\frac{ET_{\pm}}{\hbar} 
                  -\frac{\pi}{2}\,k\sigma_{\pm}\right)\nonumber\\
                  \nonumber\\
            & + & \frac{1}{\hbar\omega_0}\sum_{\pm}\sum_{k=1}^{\infty} 
                  \frac{(\pm1+2\kappa\hbar\omega_0)}
                  {2\,(1\pm\kappa\hbar\omega_0)^2}\frac{1}
                  {\sin\left[2k\pi/(1\pm\kappa\hbar\omega_0)\right]}
                  \,\sin\left(k\,\frac{ET_{\pm}}{\hbar} 
                  -\frac{\pi}{2}\,k\sigma_{\pm}\right)\nonumber\\
                  \nonumber\\
            & + & \frac{1}{\hbar\omega_0}\sum_{\pm}\sum_{k=1}^{\infty} 
                  \frac{1}{(1\pm\kappa\hbar\omega_0)^2}\frac
                  {\cos\left[2k\pi/(1\pm\kappa\hbar\omega_0)\right]}
                  {\sin^2\left[2k\pi/(1\pm\kappa\hbar\omega_0)\right]}
                  \,\cos\left(k\,\frac{ET_{\pm}}{\hbar} 
                  -\frac{\pi}{2}\,k\sigma_{\pm}\right)\nonumber\\
                  \nonumber\\
            & + & \frac{1}{\hbar\omega_0}
                  \sum_{k=1}^{\infty}
                  \frac{(-1)^{k+1}}{2\sin^2\left(k\pi\kappa\hbar\omega_0
                  \right)}
                  \,\cos\left(k\,\frac{ET_0}{\hbar} 
                  -\frac{\pi}{2}\,k\sigma_0\right),
\label{hsphtrform}
\end{eqnarray}
where the three periods $T_\pm$ and $T_0$ are given by
\bea
T_{\pm} = \frac{2\pi}{\omega_0(1\pm\kappa\hbar\omega_0)}, 
\qquad T_0 = \frac{2\pi}{\omega_0},
\label{tpm}
\eea
and the phases
\bea
\sigma_{\pm} = \frac{\pm\,2}{1\pm\kappa\hbar\omega_0}, \qquad
\sigma_{0} = -4\kappa\hbar\omega_0
\eea
play the role of non-integer Maslov integers. When added to the smooth
part \eq{smooth}, eq \eq{hsphtrform} reproduces the exact quantum spectrum
\eq{specso0}. This trace formula thus serves us as a test limit of the 
semiclassical results derived below in the non-integrable deformed cases. 
In $T_0$ we recognize the period of the classical orbits of the unperturbed 
Hamiltonian; the shifted periods $T_\pm$ have to be explained by the
periodic orbits of the perturbed system.

In the limit $\kappa\rightarrow 0$, the sum of the smooth term \eq{smooth} 
with $\omega_x$=$\omega_y$=$\omega_z$=$\omega_0$ and the oscillating term 
\eq{hsphtrform} yields the exact trace formula of the isotropic 
three-dimensional harmonic oscillator \cite{book,bj95} (which here
includes the spin degeneracy factor 2)
\bea
g(E) = \frac{1}{(\hbar\omega_0)^3}\left[E^2-\frac14(\hbar\omega_0)^2\right]
       \biggl\{ 1 + 2\,\sum_{k=1}^\infty (-1)^k\,
       \cos\left(k\frac{2\pi E}{\hbar\omega_0}\right)\biggr\}\,.
\label{g3dho}
\eea

\subsubsection{Fourier transforms}
\label{sec:fourtrans}

Since we have no explicit quantum spectra of the perturbed system and
therefore cannot derive an exact trace formula, we resort to the method
of Fourier transforms of the quantum spectrum \cite{frwi} in order to 
extract information on the periods of the system. The Hamiltonian
\eq{sysgen} has the scaling property 
\bea
\hat{H}(\eta{\bf r},\eta\hat{\bf p})
                  = \eta^2 \hat{H}({\bf r},\hat{\bf p})\,.
\label{scal}
\eea
We see below that this scaling property holds also in the classical 
limit if the SCL is used. As a consequence, the energy dependence of the 
classical dynamics and thus of the periodic orbits is simply given by
a scaling, and their actions go like $S_{po}(E)=T_{po}E$, whereby the 
periods $T_{po}=2\pi/\omega_{po}$ are energy independent (but depend on 
$\kappa$). Therefore, the peaks in the Fourier transforms of $\delta 
g(E)$ with respect to the variable $E$ will give us directly the periods 
$T_{po}$ in the time domain, whereby the peak heights are given by the 
semiclassical amplitudes ${\cal A}_{po}$ and their signs give 
information on the relative Maslov indices.

In fig \ref{fourspec} we present a series of Fourier transforms of 
$\delta g(E)$ obtained from the numerically diagonalized quantum spectra
with a coarse-graining parameter $\gamma=0.5\,\hbar\omega_0$. Shown are
here the squares of the Fourier amplitudes in the time domain, plotted
for different spin-orbit coupling strengths $\kappa$. A slightly anisotropic 
ratio of frequencies $\omega_x=\omega_0$, $\omega_y=1.1215\,\omega_0$, 
$\omega_z=1.2528\,\omega_0$ was chosen. For $\kappa=0$, the system then
has only the three isolated librating orbits along the principal axes.
Indeed, wee see at $\kappa=0$ the three dominant peaks with the
corresponding primitive periods $T_i=2\pi/\omega_i$ $(i=x,y,z)$. Their
second harmonics $(k=2)$ are also resolved; however, due to their larger
periods they are of smaller amplitude. For $\kappa>0$ this simple peak 
structure is split and ends in a completely different spectrum at 
$\kappa=0.2\,(\hbar\omega_0)^{-1}$. The dotted lines are the predictions 
from the semiclassical SCL analysis given in the next subsection; the plots 
on the right-hand side indicate the shapes of the periodic orbits for
increasing $\kappa$.

\newpage

\subsubsection{Semiclassical treatment in the SCL approach}
\label{subsec:3dscl}

We now analyze the system semiclassically in the SCL. According to sec
\ref{sec:scl}, the classical Hamiltonians to be used are 
\bea
H_\pm({\bf r},{\bf p})=\frac12\,p^2 + V({\bf r})
                       \pm{\bar\kappa}\,|{\bf C}({\bf r},{\bf p})\,|\,,
\label{3hopm}
\eea
with $\bf C$ given in \eq{sysgen} and $\bar\kappa=\hbar\kappa$ as 
discussed in sec \ref{sec:scl}. The spin-orbit term destroys the 
integrability of the harmonic oscillator, but the scaling property 
\eq{scal} is still fulfilled. Therefore the above Fourier spectra 
should give us the correct periods of the periodic orbits defined by 
the Hamiltonians \eq{3hopm}.

\Figurebb{fourspec}{-10}{70}{710}{840}{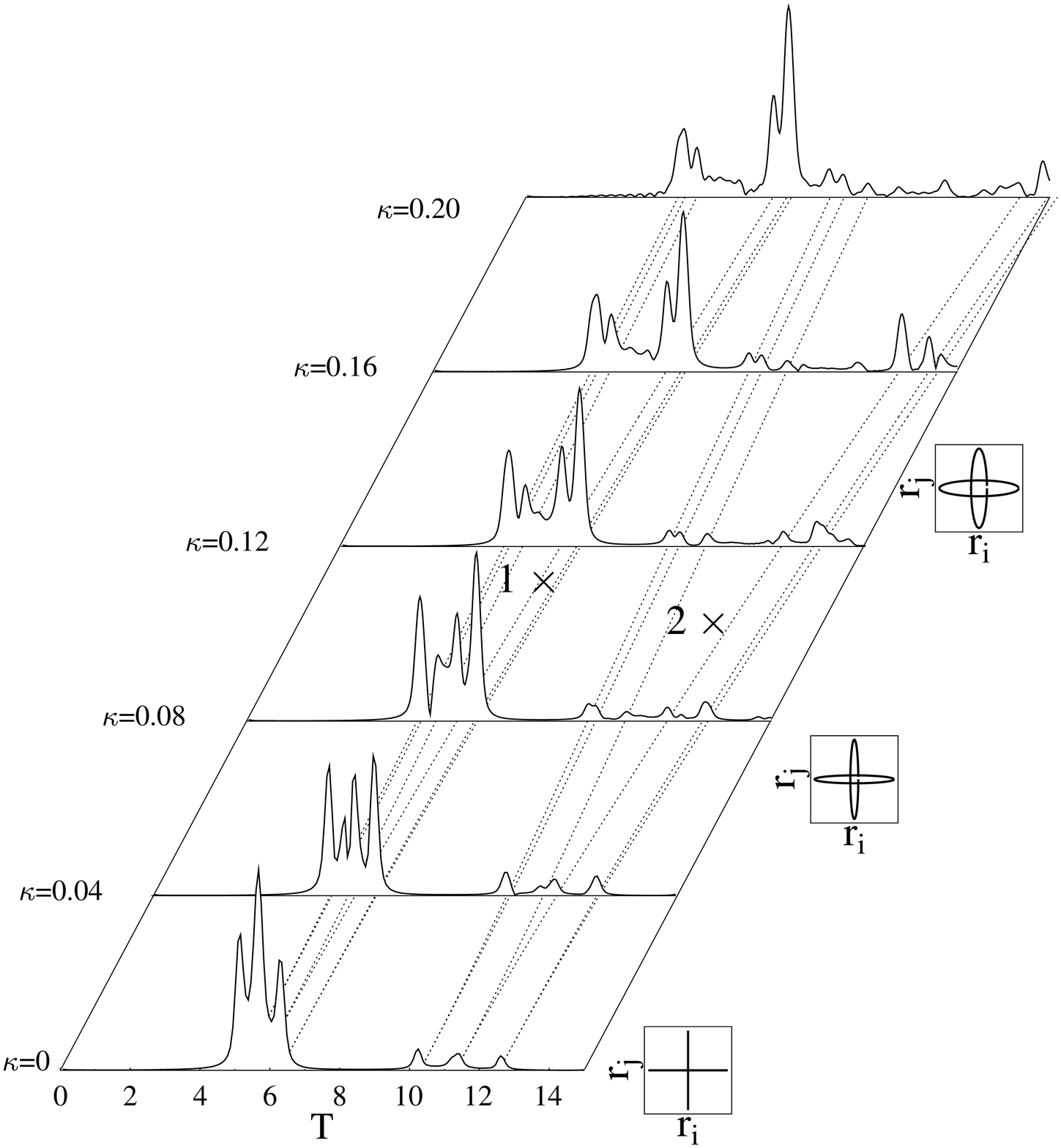}{13}{13}{
Fourier spectra of quantum-mechanical level density $\delta g(E)$ 
(coarse-grained with $\gamma=0.5\,\hbar\omega_0)$ of the 3-dimensional
harmonic oscillator with deformation $\alpha=0.1212$, $\beta=2$, and
various spin-orbit strengths $\kappa$ in units $(\hbar\omega_0)^{-1}$. 
$T$ is in units of $\omega_0^{-1}$. For the dotted lines and the 
inserts on the rhs, see fig \ref{ellipses}. 
}

\vspace*{-0.5cm}

{\bf a) Periodic orbits}

\ms

The equations of motion for the Hamiltonians \eq{3hopm} become
\begin{eqnarray}
\dot{r}_i & = & \hspace{0.32cm} p_i \hspace{0.38cm} \pm \; 
                \epsilon_{ijk}\,\bar\kappa\,\vert{\bf C}\vert^{-1}
                (C_j\omega_k^2 r_k-C_k\omega_j^2 r_j)\,,\nonumber\\
\dot{p}_i & = & -\omega_i^2 r_i \pm \;
                \epsilon_{ijk}\,\bar\kappa\,\vert{\bf C}\vert^{-1}
                (C_j\omega_i^2 p_k-C_k\omega_i^2 p_j)\,. 
                \hspace{1cm} i,j,k\in\{x,y,z\} 
\label{fulleqmot}
\end{eqnarray}
This is a non-linear coupled system of six equations, and the search for 
periodic orbits is not easy. We have determined them numerically by a 
Newton-Raphson iteration employing the stability matrix \cite{cham}. 
Special care must be taken at the MC points where ${\bf C}({\bf r,p})=0$ 
and hence the equations \eq{fulleqmot} are ill defined. In general,
this leads to discontinuities in the shapes of the periodic orbits, due 
to which their stabilities cannot be defined. We shall return to the
MC problem in sec \ref{sec:mc}. It turns out that there exist periodic
orbits which are free of MC, i.e., for which ${\bf C}({\bf r,p})$ never
becomes zero. The existence of some particularly simple orbits follows 
from the fact that the three planes $r_k=p_k=0$ $(k=x,y,z)$ in phase 
space are invariant under the Hamiltonian flow. The coupled equations of 
motion for the class of two-dimensional orbits lying in the ($i,j$) 
planes are
\begin{eqnarray}
\dot{r}_i & = & \hspace{0.32cm} p_i\hspace{0.38cm}\mp\,\epsilon_{ijk}\, 
                \bar\kappa\sign(C_k)\,\omega^2_j\, r_j\,,\nonumber\\
\dot{p}_i & = & -\omega_i^2 r_i \mp\,\epsilon_{ijk}\, 
                \bar\kappa\sign(C_k)\,\omega_i^2\, p_j\,,
                \hspace{1cm} i,j,k \in\{x,y,z\}
\label{eqmot2}
\end{eqnarray} 
where $i$ and $j$ refer to the in-plane variables and $k$ to the normal 
of each plane. Assuming that there exist solutions with $C_k\neq0$, we 
can put $\sign(C_k)=1$, since to each such orbit, there exists a 
time-reversed partner which belongs to the opposite of the two 
Hamiltonians $H_\pm$. Hence we obtain the system of equations
\begin{eqnarray}
\dot{r}_i & = & \hspace{0.32cm} p_i\hspace{0.38cm}\mp\,\epsilon_{ijk} 
                \,\bar\kappa\,\omega^2_j\, r_j\,,\nonumber\\
\dot{p}_i & = & -\omega_i^2 r_i \mp\,\epsilon_{ijk} 
                \,\bar\kappa\,\omega_i^2\, p_j\,,
                \hspace{1cm} i,j,k \in\{x,y,z\}
\label{eqmot3}
\end{eqnarray} 
which now are strictly linear and can be solved by finding the normal
modes. The solutions are two periodic orbits of ellipse form in each 
invariant plane $(i,j)$. For the two Hamiltonians $H_\pm$ this gives 
altogether twelve planar periodic orbits which come in doubly degenerate 
pairs. We therefore find six different orbits (denoted by 
$\gamma_{ij}^\pm$ in fig \ref{bif2} below) with the frequencies
\begin{eqnarray}
\omega^\pm_{ij}=\left[\left(\omega_j^2+\omega_i^2+2\,{\bar\kappa}^2
                \omega_i^2\omega_j^2 + \sqrt{\left(\omega_j^2-\omega_i^2 
                \right)^{\!2}\!+8\,{\bar\kappa}^2\omega_i^2\omega_j^2
                \left(\omega_i^2+\omega_j^2 \right) }\right)\right]^{1/2}.
\label{omegaij}
\end{eqnarray}

The rhs of fig \ref{ellipses} shows these frequencies versus the spin-orbit 
coupling parameter $\bar\kappa$. On the lhs we illustrate the periodic 
orbits for $\bar\kappa\ne0$ (ellipses) and the unperturbed libration orbits 
for $\bar\kappa=0$. The periods $T^\pm_{ij}=2\pi/\omega^\pm_{ij}$ fit 
perfectly the positions of the most pronounced peaks in the Fourier spectra 
of fig \ref{fourspec} (cf the dotted lines there), when the appropriate 
deformations $\omega_i$, $\omega_j$ are chosen. Some of the minor peaks may 
be attributed to non-planar orbits (see below).

\Figurebb{ellipses}{90}{535}{531}{680}{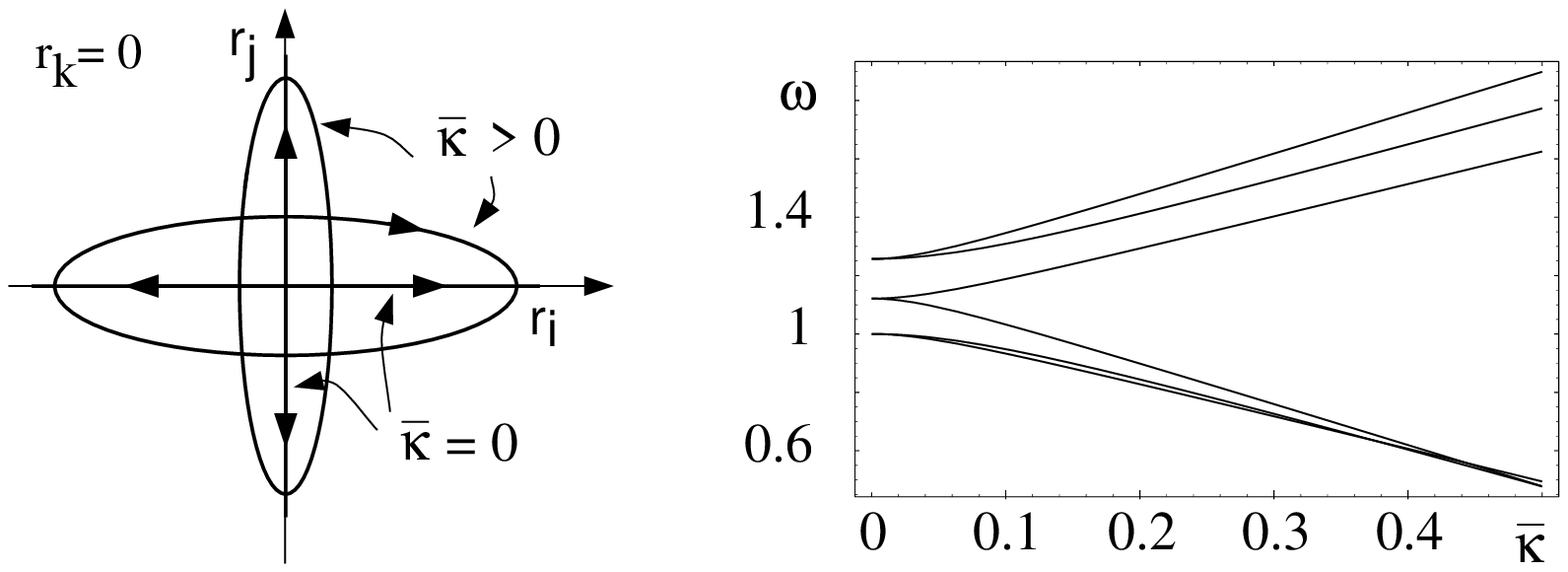}{3.1}{12}{
{\it Right:} the six frequencies $\omega^\pm_{ij}$ \eq{omegaij} 
(units: $\omega_0$) for $ij=12$, 23,\\ and 31 versus $\bar\kappa$ 
(units: $\omega_0^{-1}$). Deformations as in fig \ref{fourspec}. 
{\it Left:} sche\-matic plot of shapes of the elliptic orbits 
($\bar\kappa>0$) and the unperturbed librating orbits ($\bar\kappa=0$) 
in the $(i,j)$ plane.
}

\vspace*{-0.5cm}

In the spherical limit, the same procedure leads to a very simple 
analytical result. Due to the conserved angular momentum ${\bf L}=
{\bf r}\times{\bf p}$, most periodic orbits are planar circles. In each 
plane, the equations of motion are similar to \eq{eqmot3}, with the two 
eigenfrequencies $\omega^\pm$ and corresponding periods $T^\pm$ given by
\begin{eqnarray}
\omega^\pm=\omega_0\,(1\pm\bar\kappa \omega_0)\,, \qquad
T^\pm=\frac{2\pi}{\omega_0\,(1\pm\bar\kappa \omega_0)}\,.
\end{eqnarray}
The periods $T^\pm$ are exactly equal to the two periods $T_\pm$ in 
\eq{tpm} that appear in the exact trace formula \eq{hsphtrform} of the 
spherical system.\footnote{Note that in the present SCL approach, the 
constant $\bar\kappa$ includes a factor $\hbar$, see sec \ref{sec:scl}} 
However, the unperturbed harmonic oscillator period $T_0=2\pi/\omega_0$ 
that also appears in \eq{hsphtrform} cannot be explained by the present 
solutions in the SCL. We surmise that it might be connected to the 
existence of straight-line librating orbits; these lie, however, on mode 
conversion surfaces and cannot be treated in the present approach. In 
our ongoing studies \cite{plet} where the MC problem is avoided, we can, 
indeed, confirm the existence of periodic orbits with the period $T_0$.

Besides the above harmonic planar solutions, the full non-linear system
\eq{fulleqmot} leads also to non-planar three-dimensional periodic 
solutions with ${\bf C}({\bf r,p})>0$ (or $<0$) for which mode conversion 
does not occur. Some of these numerically obtained orbits evaluated at 
$\bar\kappa=0.2\,\omega_0^{-1}$ are shown in fig \ref{3dorb}. We also
give their periods $T_{po}$ and partial traces $\Lambda_i$ which 
determine their stabilities (see below). The orbit $\gamma_{332}^-$ has 
the period $T_{po}=10.08\,\omega_0^{-1}$ which seems to be supported 
numerically by a small peak seen in the uppermost Fourier spectrum of 
fig \ref{fourspec}. The periods of the other orbits from fig \ref{3dorb} 
could not clearly be identified in the Fourier spectra; some of these 
orbits are too unstable and some of the periods lie too close to those 
of the leading planar orbits.

\Figurebb{3dorb}{0}{-10}{1560}{1675}{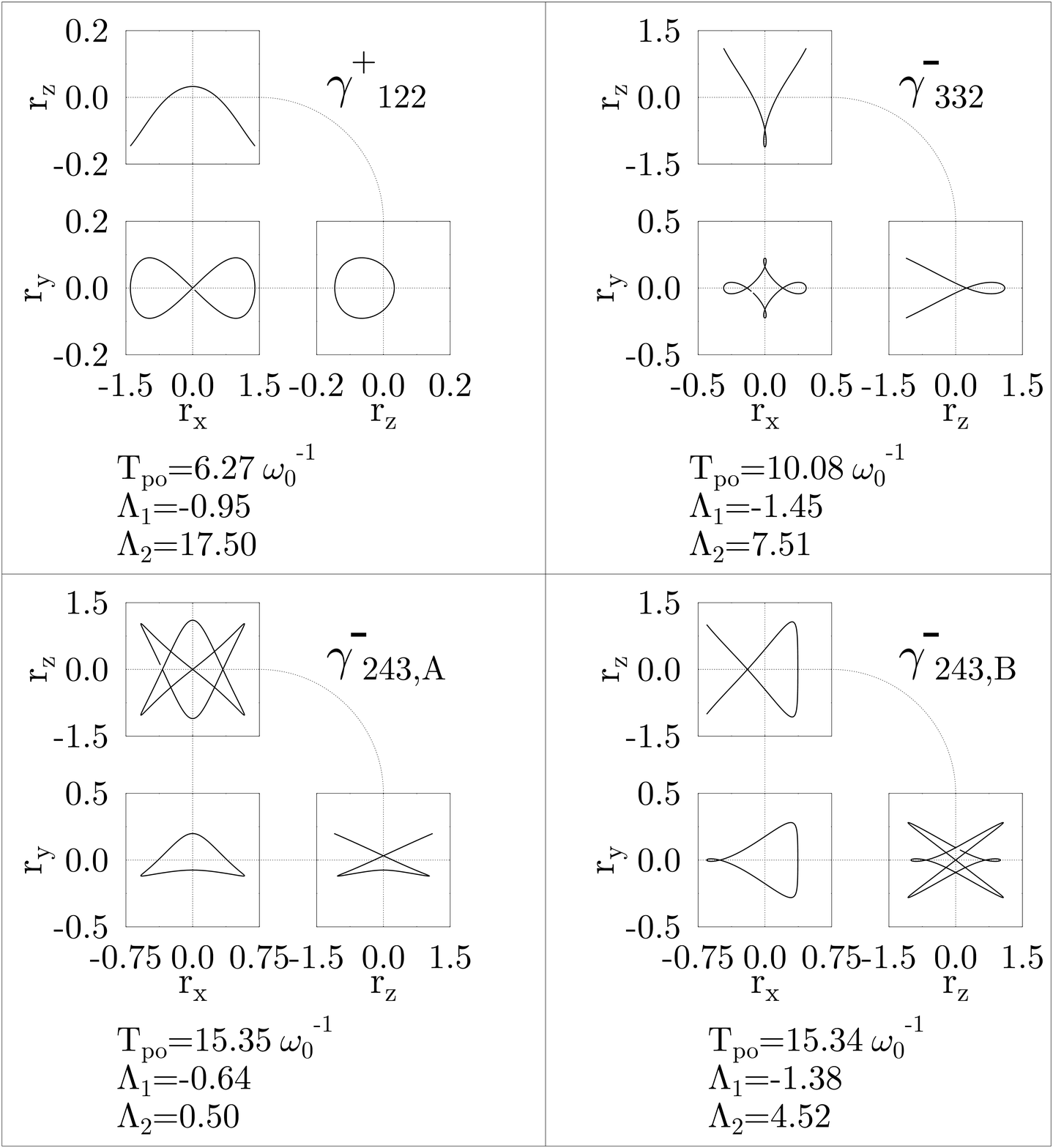}{10.7}{15}{
Shapes of four 3-dimensional non-planar orbits found in the harmonic 
oscillator with deformations as in fig \ref{fourspec} and spin-orbit 
interaction $\bar\kappa=0.2\,\omega_0^{-1}$, projected onto the three 
spatial planes. (See text for the periods $T_{po}$ and stability 
traces $\Lambda_1$, $\Lambda_2$.)
}  

\newpage

{\bf Stability amplitudes and trace formula for isolated orbits}

\ms

The amplitude ${\cal A}_{po}$ of a periodic orbit in the trace formula 
is strongly dependent on its stability. For isolated orbits in a 
two-dimensional system, the factor $|\det(\widetilde{\M}_{po}-\unit)|$ 
in \eq{amppo} equals $|2-\hbox{tr}\,\widetilde{\M}_{po}|=
|2-(\lambda_1+\lambda_2)|=|2-(\lambda_1+1/\lambda_1)|$, where 
$\lambda_i$ ($i=1,2$) are the eigenvalues of the stability matrix 
$\widetilde{\M}_{po}$, and thus the quantity tr$\,\widetilde{\M}_{po}$
contains all information about the stability of an orbit. For a system 
in $d\geq2$ dimensions, we can write
\begin{eqnarray}
|\det(\widetilde{\M}_{po}-\unit)| = 
                       \prod\limits_{i=1}^{d-1}|\Lambda_i-2|\,,\qquad
\label{stabd}
\end{eqnarray}
where the ``partial traces'' $\Lambda_i$ are the sums of pairs 
$\lambda_i$, $1/\lambda_i$ of mutually inverse eigenvalues of the 
$(2d-2)\,$ dimensional stability matrix $\widetilde{\M}_{po}$: 
$\Lambda_i=\lambda_i+1/\lambda_i$ ($i=1,2,\dots,d$$-$1). An orbit is 
stable when $|\Lambda_i|<2$ for all $i$, unstable when $|\Lambda_i|>2$ 
for all $i$, and mixed stable (or loxodromic) in all other cases. For 
the latter cases, the stability depends on the phase-space direction of 
a perturbation. Whenever $\Lambda_i=+2$ for any partial trace, a 
bifurcation occurs and the stability denominator \eq{stabd} becomes zero. 
In such a situation one has to resort to uniform approximations 
\cite{ozha,ssun} in order to obtain finite semiclassical amplitudes. 
Non-isolated periodic orbits with $\Lambda_i=2$ occur in degenerate 
families for systems with continuous symmetries and are characteristic of 
integrable systems; for these, the amplitudes must be obtained differently 
\cite{babl,stru,beta,crli}. The symmetry breaking away from integrability 
can also be handled perturbatively \cite{crpe} or with suitable uniform 
approximations \cite{tgu,hhun}. 

Most of the orbits that we have found, both planar and non-planar, 
undergo bifurcations when the spin-orbit parameter $\bar\kappa$ or the 
deformation parameters $\alpha$, $\beta$ are varied. A typical scenario 
is illustrated in fig \ref{bif2}, where $\alpha$ is varied at fixed 
$\bar\kappa=0.1\,\omega_0^{-1}$ and $\beta=2$. Shown are the
partial traces $\Lambda_i$ of the involved orbits. One of the planar 
ellipse orbits ($\gamma^-_{\rm xy}$) lying in the $(x,y)$ plane undergoes 
an isochronous pitchfork bifurcation at $\alpha=0.3977$. The new-born 
pair of orbits ($\gamma^-_{\rm bif}$) are degenerate with respect to a 
reflection at the $(x,y)$ plane. They are non-planar warped ellipses 
which rotate out of the $(x,y)$ plane when $\alpha$ is increased, and 
then approach the $(x,z)$ plane. Through an inverse pitchfork bifurcation 
at $\alpha=0.4450$, they finally merge with another planar ellipse orbit 
($\gamma^-_{\rm zx}$) lying in the $(x,z)$ plane. Near $\alpha\sim 0.425$,
the orbit $\gamma^-_{\rm bif}$ suffers from two more bifurcations (the
other orbits involved thereby are not shown).

\Figurebb{bif2}{10}{20}{709}{480}{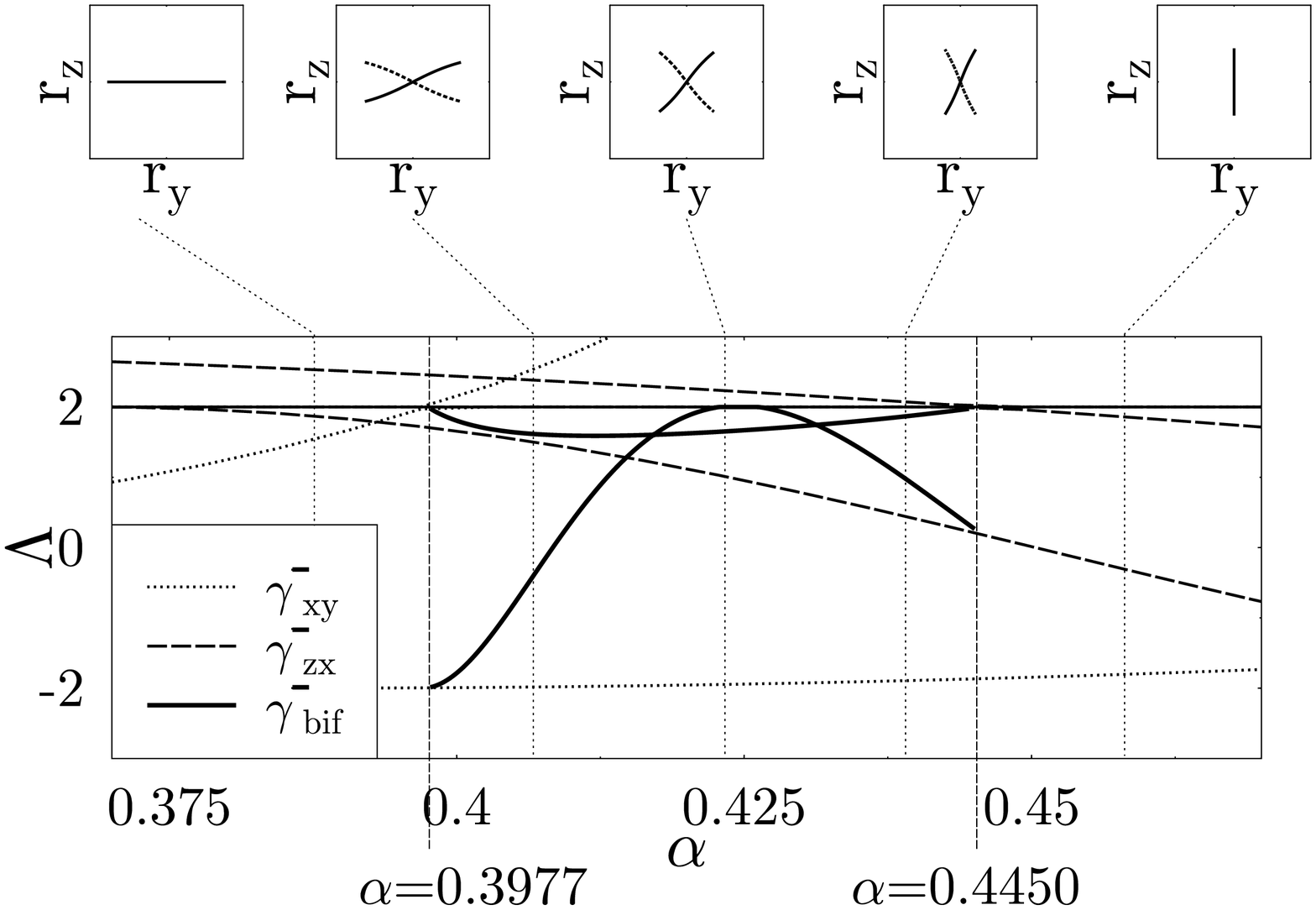}{6.3}{15}{
Bifurcation of a planar ellipse orbit under variation of the deformation 
parameter $\alpha$. {\it Upper panels:} projection of the orbits onto
the $(y,z)$ plane. {\it Lower panel:} partial traces $\Lambda_i$ of the
involved orbits versus $\alpha$ (see text for details).
}

This example shows that the classical dynamics of the Hamiltonians $H_\pm$
is mixed and quite complicated due to the unavoidable bifurcations. In
principle, isolated bifurcations can be handled using the by now well-known
uniform approximations \cite{ozha,ssun}. These fail, however, when two
bifurcations lie so close that the difference between the corresponding
actions $S_{po}$ becomes comparable to or less than $\hbar$. It would lead
outside the scope of the present study to attempt to regulate the Gutzwiller
amplitudes by uniform approximations. In fig \ref{bif1} we show by crosses
the critical values of the frequencies $\omega_y$ and $\omega_z$ (in
units of $\omega_x=\omega_0$) versus $\bar\kappa$, at which bifurcations of 
the planar orbits $\gamma_{ij}^\pm$ occur for fixed values of the 
deformation parameter $\beta=2$ (left) and $\beta=3$ (right). The other 
deformation parameter $\alpha$ is given via \eq{defpar}. In the deformation 
regions below the dashed lines, where no bifurcations occur, the 
semiclassical amplitudes can be used without further uniform approximation.

\Figurebb{bif1}{-20}{40}{665}{280}{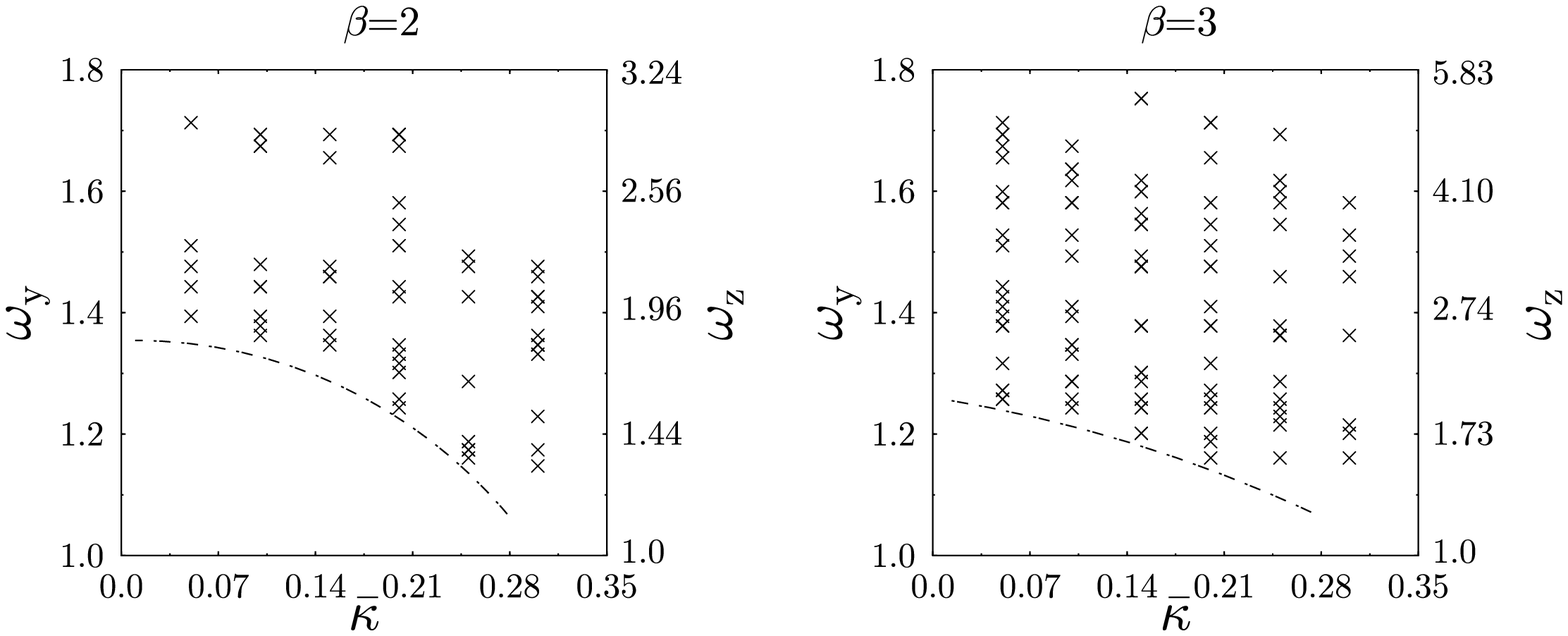}{5.8}{16.5}{
Bifurcations of the planar periodic orbits $\gamma_{ij}^\pm$ in the 
3-dimensional harmonic oscillator with spin-orbit interaction, given
by the SCL Hamiltonians $H_\pm$ in \eq{3hopm}. The crosses give the 
critical values of the frequencies $\omega_y$ and $\omega_z$ versus 
$\bar\kappa$ for $\beta=2$ (left) and $\beta=3$ (right). No bifurcations 
were found in the regions below the dashed lines.
}

\vspace*{-0.5cm}

For the contributions of all the isolated periodic orbits away from the
bifurcations, we therefore use the trace formula
\begin{eqnarray}
\delta g_\gamma(E) = \frac{1}{\hbar\pi}\sum_{po} 
                     e^{-(\gamma T_{po}/2\hbar)^2}
                     \frac{T_{ppo}}{\sqrt{\prod_{i=1}^{d-1}|\Lambda_i-2|}}\,
                     \cos\left(\frac{1}{\hbar}\,S_{po} 
                     +\Delta\Phi_{po}
                     -\frac{\pi}{2}\,\sigma_{po} \right),
\label{trace3d}
\end{eqnarray}
whereby the sum $po$ explicitly includes all periodic orbits of both 
Hamiltonians $H_\pm$. The Maslov indices $\sigma_{po}$ were evaluated 
with the methods developed by Creagh {\it et al}~ \cite{crol}, employing 
the recipes given in appendix D of \cite{book}. The terms $\Delta\Phi_{po}$ 
are the phase corrections \eq{deltaS}. For the planar ellipse orbits 
lying in the $(i,j)$ planes, we find $\lambda_\pm^{B}=0$ and 
$\lambda_\pm^{NN}=-\bar\kappa\,\omega_k^2/2$, so that $\Delta\Phi_{po}
=-\bar\kappa\,\pi\omega_k^2/\omega^{\pm}_{ij}$. 

In fig \ref{dge} we show the results obtained for the situation 
$\alpha=0.1212$, $\beta=2$, $\bar\kappa=0.1\,\omega_0^{-1}$, for 
which no close-lying bifurcations exist.
The quantum-mechanical coarse-grained level density $\delta g(E)$ is
shown by the solid lines (QM) and includes the spin-orbit interaction
in both curves a) and b). The semiclassical results (SC) are shown 
by dashed lines; in a) without spin-orbit interaction, which again 
demonstrates that the latter dramatically changes the level density; 
in b) with spin-orbit interaction through the trace formula \eq{trace3d}.
Only the six primitive planar orbits have been used. We see that this 
already leads to an excellent agreement with quantum mechanics, except at
very low energies where semiclassics usually cannot be expected to work.
The curve SC in the lowest panel c) shows the semiclassical result over 
a larger energy scale.

Similar results were obtained in the region of deformations and
$\bar\kappa$ values below the dashed lines in fig \ref{bif1}, where
bifurcations do not occur. In all these cases, it turned out that the 
inclusion of the six primitive planar orbits was sufficient within the 
resolution given by the coarse-graining width $\gamma=0.5\,\hbar\omega_0$. 
This result is in agreement with the Fourier analysis of sec 
\ref{sec:fourtrans} of the quantum spectra, where all dominant peaks 
correspond to the periods of these six orbits.

\Figurebb{dge}{0}{30}{1160}{1290}{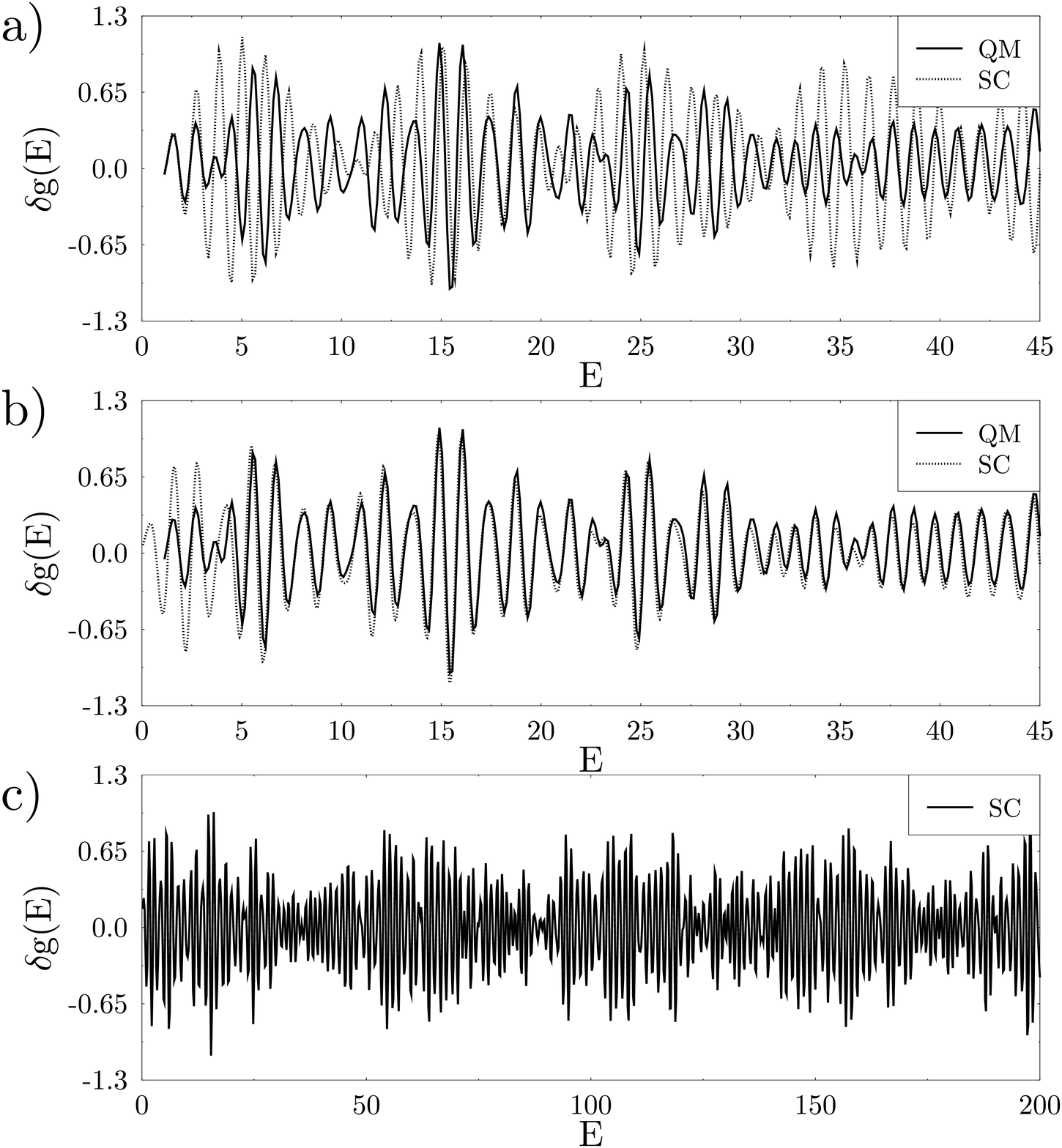}{14}{15}{
Coarse-grained level density $\delta g(E)$ of 3-dimensional harmonic
oscillator with spin-orbit interaction $\bar\kappa=0.1\,\omega_0^{-1}$ 
(other parameters as in fig \ref{fourspec}). 
{\it Upper panels:} solid lines (QM) give the quantum-mechanical 
results with spin-orbit interaction. Dashed lines (SC) give the 
semiclassical results according to \eq{trace3d}, calculated a) 
without and b) with spin-orbit interaction. {\it Bottom panel:} c) 
same as SC in b), but over a larger energy region. 
}

\newpage

\section{The problem of mode conversion}
\label{sec:mc}

In this section we want to discuss the mode conversion (MC) problem that
arises in the strong-coupling limit (SCL) following \cite{lif1,lif2}. In
particular, we will discuss an intuitive method, suggested by Frisk and
Guhr \cite{frgu}, to partially avoid the MC problem. This method can
qualitatively explain some of the peaks observed in the Fourier spectra 
of the quantum-mechanical level densities $\delta g(E)$. It can, however,
not be used to calculate the amplitudes ${\cal A}_{po}$ required for the
semiclassical trace formula.

To this purpose, we return to the 2DEG with lateral harmonic confinement 
discussed in sec \ref{sec:2dho}, but without external magnetic field 
($B_0=0$). We shall again assume the oscillator frequencies $\omega_x$
and $\omega_y$ to be incommensurable (i.e., $\omega_x/\omega_y$ is 
irrational). Then, the system without spin-orbit interaction has only  
the isolated self-retracing librating orbits along the axes, and the
weak-coupling limit (WCL) approach of \cite{boke} cannot handle the
spin-orbit interaction (except for a trivial spin factor 2 in the
trace formula). We therefore have to resort to the SCL approach. In 
order to simplify the discussion and focus on the important points,
we set $m^*=1$ and ignore the diagonal elements of the spin-orbit 
interaction. We thus start from the Hamiltonian
\begin{eqnarray}
\widehat{H} = \frac12\left(\,\hat{p}_x^2+\hat{p}_y^2+ 
              \omega_x^2 r_x^2+\omega^2_y r_y^2\,\right)\unit 
            + \hbar\kappa\left(\! \begin{array}{cc}
                  \!0 & \!\!-\hat{p}_x\!-\!\i \hat{p}_y \\      
                  \hat{p}_x\!+\!\i \hat{p}_y & ~0  
                               \end{array}\!\!\right),
\label{mcham}
\end{eqnarray}
which in the SCL leads to the classical Hamiltonians 
($\bar\kappa=\hbar\kappa$)
\bea
H_\pm = \frac12\left(p_x^2+p_y^2+\omega_x^2 r_x^2+\omega^2_y r_y^2\right) 
        \pm \bar\kappa\sqrt{p_x^2+p_y^2}\,.
\label{mchamcl}
\eea

Before taking the semiclassical limit, we first perform a Fourier 
analysis of the quantum spectrum of \eq{mcham} which is easily
diagonalized in the unperturbed harmonic-oscillator basis. The
Hamiltonian \eq{mcham} does not possess the scaling property \eq{scal}
of the three-dimensional system studied in sec \ref{sec:3dho}.
However, we can use the method of \cite{frwi} by scaling the parameter
$\bar\kappa$ away. Dividing equation \eq{mchamcl} by ${\bar\kappa}^2$ 
and introducing the scaled variables $\tilde {r}_i = {r}_i/\bar\kappa$,
$\tilde {p}_i = {p}_i/\bar\kappa$, we obtain the scaled Hamiltonians
\begin{eqnarray}
\widetilde{H}_\pm = \frac12 \left(\tilde{p}_x^2+\tilde{p}_y^2+ 
                    \omega_x^2\tilde{r}_x^2+\omega^2_y\tilde{r}_y^2\right) 
                    \pm\sqrt{\tilde{p}_x^2+\tilde{p}_y^2}
                  = E/{\bar\kappa}^2 = e\,,
\label{hpmscal}
\end{eqnarray}
so that the classical dynamics does not depend explicitly on $\bar\kappa$
but is determined only by the scaled energy variables $e$. Therefore,
a Fourier transform of the quantum spectra along the path in the 
($E,\kappa$) plane with constant $E/\kappa^2$ leads in the time domain
to the quasiperiods ${\widetilde T}^\pm_{po}=s^\pm_{po}(e)/e$ of the 
periodic orbits of the Hamiltonians \eq{hpmscal}, whereby $s^\pm_{po}$ 
are their scaled actions.
 
In the lower part of fig \ref{mocofour} below we show the result of 
the Fourier transforms, taken at two different scaled energies. For 
$e=10^5\hbar\omega_0/{\bar\kappa}^2$ (dashed line) only two peaks are 
seen. For $e=30\,\hbar\omega_0/{\bar\kappa}^2$ (solid line), these are 
slightly shifted and remain the dominant peaks, whereas four additional 
small peaks appear (the second of these extra peak is almost absorbed 
in the left dominant peak).

We now analyze the classical dynamics of the Hamiltonians \eq{hpmscal}.
Like in the three-dimensional system analyzed in sec \ref{subsec:3dscl},
we can find orbits lying in the invariant subspaces of phase space with
$\tilde{r}_i=\tilde{p}_i=0$ for one of the degrees of freedom $i$ ($x$ or 
$y$). For the other degree, the one-dimensional equations of motion become 
\begin{eqnarray}
\dot{\tilde r}_i & = & \tilde{p}_i \pm \sign \tilde{p}_i\,,\nonumber \\
\dot{\tilde p}_i & = & -\omega_i^2\, \tilde{r}_i\,.
\label{eom1d}
\end{eqnarray}
On the axes $\tilde{p}_i=0$ in phase space these equations are ill 
defined. It is still possible to solve the equations for ${\tilde p}_i 
\ne 0$, which leads to portions of a circle in the $({\tilde p}_i,
{\tilde r}_i)$ plane for each sign of ${\tilde p}_i$. One may then 
connect these partial trajectories to form periodic orbits whose shapes 
have, however, kinks. This is illustrated in the upper part a) of fig 
\ref{phasespace}. Due to the kinks, these orbits are not differentiable 
and their stabilities cannot be defined. Their periods can, however, be 
calculated analytically and become
\begin{eqnarray}
T^{\mathrm{adia}}_{i,+}(e)
  = \frac{2}{\omega_i} \arccos\left( \frac{1-2e}{1+2e} \right),
    \qquad T^{\mathrm{adia}}_{i,-}(e)
  = \frac{4\pi}{\omega_i} 
  - \frac{2}{\omega_i} \arccos\left( \frac{1-2e}{1+2e} \right).
\end{eqnarray}
From the area enclosed by the orbits, we can also obtain the scaled 
actions:
\begin{eqnarray}
s^{\mathrm{adia}}_{i,\pm}(e) = \left(e-\frac{1}{2}\right) 
 T^{\mathrm{adia}}_{i,\pm}(e)-\frac{2}{\omega_i}\sqrt{2e}\,. 
\label{actadia}
\end{eqnarray}
We use the superscript ``adia'' because the Hamiltonians $H_\pm$
correspond to the adiabatic situation where the spin polarizations
are fixed.

\Figurebb{phasespace}{0}{220}{496}{690}{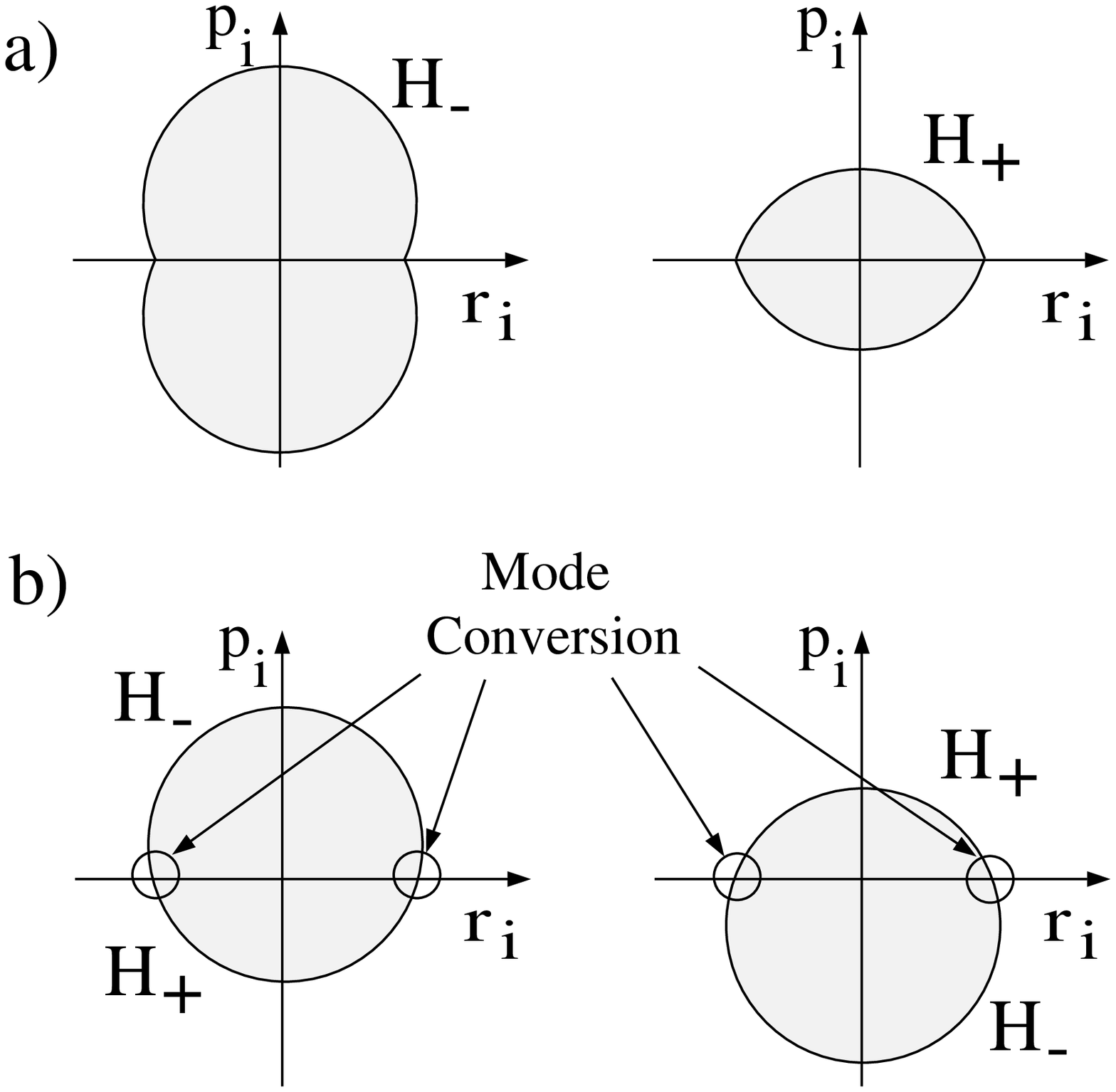}{6.5}{11}{
{\it Upper part:} {\bf a)} Periodic orbits found from the adiabatic 
Hamiltonians \eq{hpmscal}. {\it Lower part:} {\bf b)} Diabatic periodic 
orbits found by enforcing spin flips $H_+ \leftrightarrow H_-$ at the 
mode conversion points.
}

\vspace*{-0.5cm}

An alternative way to use the partial solutions found from \eq{eom1d}
has been proposed by Frisk and Guhr \cite{frgu}: instead of joining
the two portions obtained for both signs of ${\tilde p}_i$ on one and
the same of the Hamiltonians $H_\pm$, one switches between the $H_\pm$, 
enforcing a spin flip at the MC points: $H_+\longleftrightarrow H_-$. 
This corresponds to the transition from the adiabatic to a diabatic 
basis. The orbits thus obtained are continuous circles with continuous 
derivatives, as illustrated in the lower part b) of fig \ref{phasespace},
and correspond to simple harmonic librations along the $i$ axes. Their 
periods and actions are easily found to be
\begin{eqnarray} 
T^{\mathrm{dia}}_i = \frac{2\pi}{\omega_i}\,, \qquad
s^{\mathrm{dia}}_i(e) = \left(e+\frac{1}{2}\right) 
                                  \frac{2\pi}{\omega_i}\,.
\label{actdia}
\end{eqnarray}
The superscript ``dia'' indicates that we call these the diabatic
orbits. Their periods are those of the Hamiltonian without spin-orbit
coupling: $T^{\mathrm{dia}}_i = T^{(0)}_i=2\pi/\omega_i$. Although 
their shapes are continuous and differentiable, their stabilities 
still cannot be calculated, because the Hessian matrices of the $H_\pm$ 
in phase space are singular at the MC points. In the limit 
$e\rightarrow\infty$, the
pairs of adiabatic orbits merge into the diabatic orbits, and we get
\bea
T^{\mathrm{adia}}_{i,\pm}(e) \longrightarrow T^{\mathrm{dia}}_i\,,
\qquad
\frac{s^{\mathrm{adia}}_{i,\pm}(e)}{e} \longrightarrow 
\frac{s^{\mathrm{dia}}_i(e)}{e} \longrightarrow  T^{\mathrm{dia}}_i\,.
\label{dialim}
\eea

\Figurebb{mocofour}{0}{10}{908}{820}{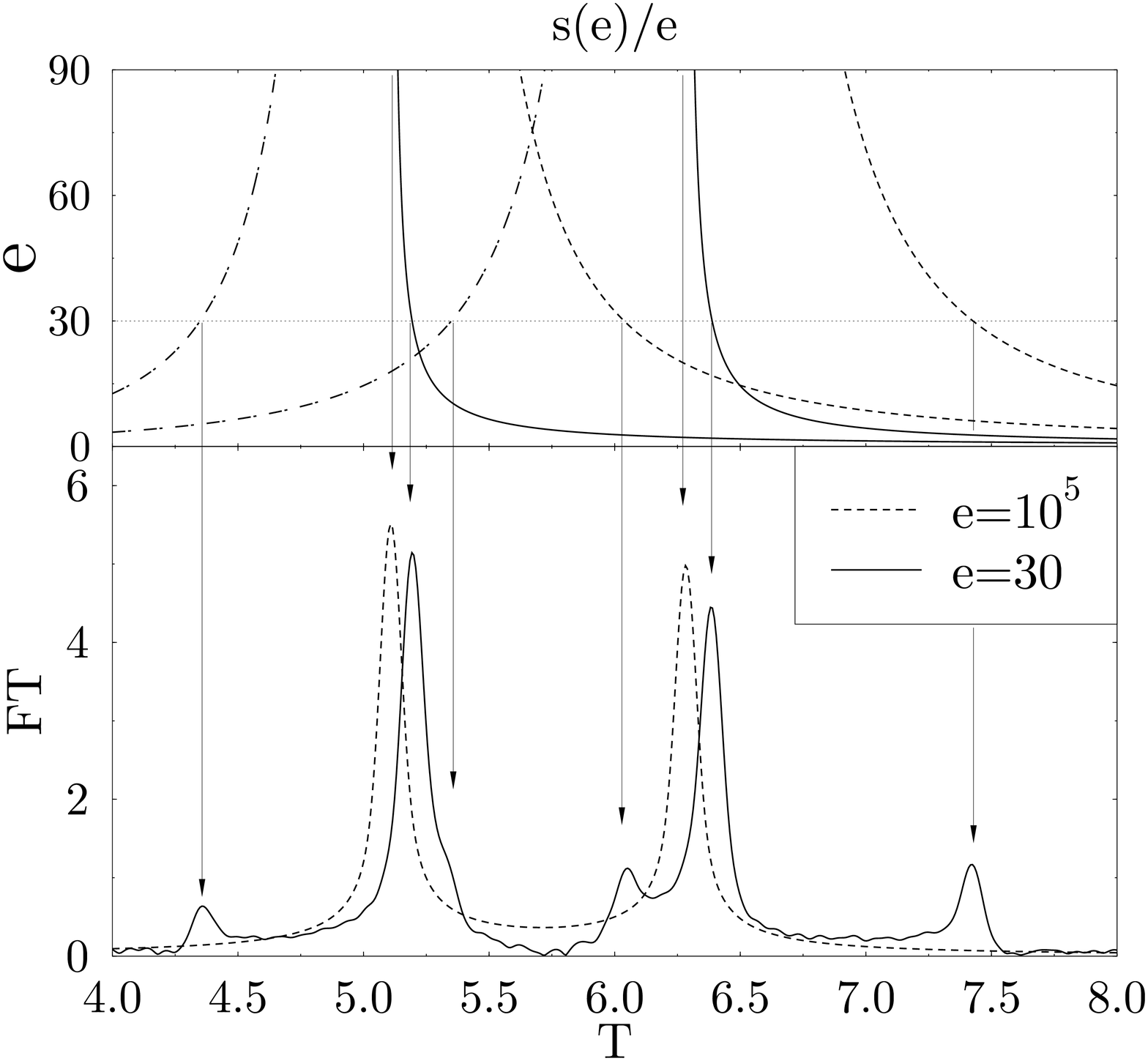}{8}{13.5}{
{\it Lower part:} Fourier spectra like in fig \ref{fourspec} of the
quantum spec\-trum of the Hamiltonian \eq{mcham} with $\gamma=0.2\,
\hbar\omega_0$, evaluated for two values of the scaled energy $e$ (units: 
$\hbar\omega_0/{\bar\kappa}^2$). Deformation: $\omega_x=\omega_0$,
$\omega_y=1.23\,\omega_0$. {\it Upper part:} quasiperiods $s(e)/e$ from
\eq{actadia} and \eq{actdia} for adiabatic orbits (dashed lines) and 
diabatic orbits (solid lines), respectively, of Hamiltonians \eq{hpmscal} 
versus scaled energy $e$.
}

\vspace*{-0.5cm}

In the upper part of fig \ref{mocofour}, we show the curves 
$s^{\mathrm{adia}}_{i,\pm}(e)/e$ by the short and long dashed lines,
and the curves $s^{\mathrm{dia}}_{i}(e)/e$ by the solid lines. We see
that their values at $e=30\,\hbar\omega_0/{\bar\kappa}^2$ correspond
exactly to the six peaks appearing in the corresponding Fourier
spectrum. The two dominant peaks correspond to the diabatic orbits, 
and the four small peaks correspond to the adiabatic orbits. For
$e=10^5\hbar\omega_0/{\bar\kappa}^2$, the only two peaks correspond
to the asymptotic values of $s^{\mathrm{dia}}_{i}(e)/e=T^{\rm dia}_i$,
in agreement with the limit \eq{dialim}.

The evidence of diabatic orbits according to the above spin-flip
hypothesis had already been observed by Frisk and Guhr \cite{frgu}.
They did, however, not recognize any signatures in their Fourier 
spectra corresponding to periods of adiabatic orbits involved with MC 
points, such as we have found them in the four minor peaks of fig
\ref{mocofour}. Their conclusion was therefore that spin-flips always 
occur at the MC points. Our results seem to suggest that both kind of 
dynamics occur. The dominant Fourier peaks are, indeed, those 
corresponding to the diabatic orbits which undergo spin flips at the MC 
points. However, there must also exist a finite probability that the 
orbits stay on the adiabatic surfaces $H\pm=E$, leading to the smaller 
peaks positioned at the correct adiabatic quasiperiods $s(e)/e$.

The semiclassical amplitudes required for the trace formula cannot
be calculated for the present system, neither using diabatic nor 
adiabatic orbits. The fully polarized treatment of the spin variables 
used in the SCL approach is obviously not flexible enough to account 
for the full dynamics, although the Fourier analysis of the quantum
spectra suggest that there is some partial truth to it. A more
complete semiclassical description of the spin motion should allow 
for a balanced mixture of adiabatic and diabatic spin motion.

\newpage

\section{Summary and conclusions}

We have derived semiclassical trace formulae for several nonrelativistic
two- and three-dimensional fermion systems with spin-orbit interactions 
of Rashba and Thomas type. We have thereby employed the weak-coupling 
limit (WCL) developed by Bolte and Keppeler \cite{boke}, and the 
strong-coupling limit (SCL) of Littlejohn and Flynn \cite{lif1} with
extensions and justifications of \cite{frgu,boke}. 

In the WCL approach, the spatial motion of the particles is taken into
account only using the periodic orbits of the system without spin-orbit
interaction. The spin motion is included adiabatically via the trace of 
a spin transport matrix {\tt\large d}$(t)$ which describes the spin 
precession about the instantaneous magnetic field provided by the 
spin-orbit interaction. Nevertheless, we found that for a 2DEG in an 
external magnetic field with and without lateral anharmonic confinement, 
the WCL yields excellent results. In the free case, for which an exact 
trace formula can be derived, the semiclassical WCL reproduces the exact
leading-order terms both in $\hbar$ and in the spin-orbit coupling
constant $\kappa$. In the laterally confined case, the gross-shell 
structure of
the coarse-grained quantum level density was very accurately reproduced
numerically. From our results it can be seen that in the limit of very 
large spin-orbit constants $\kappa$, the missing higher-order terms may 
restrict the 
applicability of this method. A particular situation, where the WCL 
approach misses the effects of the spin-orbit interaction totally, is that
where only self-retracing isolated periodic orbits exist, for which the
trace of {\tt\large d}$(T_{po})$ 
only yields a trivial spin degeneracy factor 2.

We have studied the SCL approach for two systems possessing exclusively 
self-retracing isolated orbits for which the WCL approach fails, namely
two- and three-dimensional harmonic oscillators with irrational
frequency ratios. In the SCL approach the mode conversion (MC) problem,
arising at points in phase space where the spin-orbit interaction
locally is zero, imposes severe restrictions, since singularities in the
equations of motion and the linear stability analysis of periodic orbits
arise at the MC points. However, in 
the three-dimensional case, which provides a realistic shell model for 
light atomic nuclei, we found that the leading orbits with 
shortest periods are free of MC and lead to excellent results of the 
semiclassical trace formula for the coarse-grained level density, as 
long as bifurcations of these orbits are avoided. (The latter, when 
they are sufficiently separated in phase space, can be taken into 
account using well-developed uniform approximations and do, in 
principle, not affect the applicability of the SCL approach.) 

In the two-dimensional model of an anisotropic quantum dot, the MC 
problem could not be avoided. By a Fourier analysis of the quantum 
spectrum we have provided some support of the diabatic spin-flip 
hypothesis put forward by Frisk and Guhr \cite{frgu}. We have to 
extend their conclusions, though, in the sense that there is evidence 
for a mixture of both diabatic and adiabatic classical motion on the 
two spin-polarized energy surfaces $H_\pm=E$, whereby the diabatic periods 
dominate the Fourier spectra. But even in the diabatic spin-flip limit, 
the semiclassical amplitudes of the trace formula cannot be calculated, 
and the MC problem therefore remains essentially unsolved. The 
connection between the existence of MC points and real spin-flip 
processes should therefore be taken with caution. In order to study
the physical relevance of spin flips in the presence of a spin-orbit 
interaction, there is a definite need for a better analytical
semiclassical treatment of the spin degrees of motion that is free
of singularities and allows for a balanced mixture of adiabatic and 
diabatic spin motion. A new approach \cite{plet} in which the MC problem 
does not arise is presently being developed and will be presented in more 
detail in a forthcoming paper. 

\bs
\bs

\noindent
{\Large \bf Acknowledgements}

\bs

\noindent
We are grateful to J. Bolte, S. Keppeler, M. Langenbuch, M. Mehta, M. 
Pletyukhov, O. Zaitsev and U. R\"ossler for stimulating discussions. This 
work has been supported by the Deutsche Forschungsgemeinschaft.

\end{document}